\documentclass[psfig]{aa}
\input psfig.tex

\twocolumn
                                                                               
\begin{document}

   \thesaurus{06         
              (03.11.1)}  

   \title{The ROSAT-ESO Flux-Limited X-Ray (REFLEX) Galaxy Cluster
         Survey\,III: The Power Spectrum\thanks{Based
         partially on observations collected at the European Southern
         Observatory La\,Silla, Chile} }

   \titlerunning{The REFLEX Power Spectrum}

   \author{Peter Schuecker\,$^{(1)}$, 
           Hans B\"ohringer\,$^{(1)}$,  
           Luigi Guzzo\,$^{(2)}$, 
           Chris A. Collins\,$^{(3)}$, 
           Doris M. Neumann\,$^{(4)}$, 
           Sabine Schindler\,$^{(3)}$, 
           Wolfgang Voges\,$^{(1)}$,  
	   Sabrina De\,Grandi\,$^{(2)}$,   
           Guido Chincarini\,$^{(2,5)}$,             
           Ray Cruddace\,$^{(6)}$,  
           Volker M\"uller\,$^{(7)}$,  
	   Thomas H. Reiprich\,$^{(1)}$, 
           J\"org Retzlaff\,$^{(1)}$ and 
	   Peter Shaver\,$^{(8)}$ }
                                       
   \authorrunning{Schuecker et al.}

   \offprints{Peter Schuecker\\ peters@mpe.mpg.de} 

   \institute{
    $^{(1)}$ Max-Planck-Institut f\"ur extraterrestrische Physik,
             Garching, Germany\\
    $^{(2)}$ Osservatorio Astronomico di Brera, Merate, Italy\\
    $^{(3)}$ Liverpool John Moores University, Liverpool, U.K.\\
    $^{(4)}$ CEA Saclay, Service d'Astrophysique, Gif-sur-Yvette,
             France\\
    $^{(5)}$ Dipartimento di Fisica, Universita degli Studi di Milano,
             Italy\\
    $^{(6)}$ Naval Research Laboratory, Washington D.C., U.S.A.\\
    $^{(7)}$ Astrophysikalisches Institut, Potsdam, Germany \\
    $^{(8)}$ European Southern Observatory, Garching, Germany.}

   \date{Received ..... ; accepted .....}                         
   
   \maketitle   
                                                                             
   \markboth {The REFLEX Power Spectrum}{}


\begin{abstract} 
We present a measure of the power spectrum on scales from $15$ to
$800\,h^{-1}\,{\rm Mpc}$ using the ROSAT-ESO Flux-Limited X-Ray
(REFLEX) galaxy cluster catalogue. The REFLEX survey provides a sample
of the 452 X-ray brightest southern clusters of galaxies with the
nominal flux limit $S=3.0\times 10^{-12}\,{\rm erg}\,{\rm
s}^{-1}\,{\rm cm}^{-2}$ for the ROSAT energy band
$(0.1-2.4)$\,keV. Several tests are performed showing no significant
incompletenesses of the REFLEX clusters with X-ray luminosities
brighter than $10^{43}\,{\rm erg}\,{\rm s}^{-1}$ up to scales of about
$800\,h^{-1}\,{\rm Mpc}$. They also indicate that cosmic variance
might be more important than previous studies suggest. We regard this
as a warning not to draw general cosmological conclusions from cluster
samples with a size smaller than REFLEX. Power spectra, $P(k)$, of
comoving cluster number densities are estimated for flux- and
volume-limited subsamples. The most important result is the detection
of a broad maximum within the comoving wavenumber range $0.022\le k\le
0.030\,h\,{\rm Mpc}^{-1}$. The data suggest an increase of the power
spectral amplitude with X-ray luminosity. Compared to optically
selected cluster samples the REFLEX $P(k)$ is flatter for wavenumbers
$k\le 0.05\,h\,{\rm Mpc}^{-1}$ thus shifting the maximum of $P(k)$ to
larger scales. The smooth maximum is not consistent with the narrow
peak detected at $k=0.05\,h\,{\rm Mpc}^{-1}$ using the Abell/ACO
richness $\ge 0$ data. In the range $0.02\le k \le 0.4\,h\,{\rm
Mpc}^{-1}$ general agreement is found between the slope of the REFLEX
$P(k)$ and those obtained with optically selected galaxies. A
semi-analytic description of the biased nonlinear power spectrum in
redshift space gives the best agreement for low-density Cold Dark
Matter models with or without a cosmological constant.
\end{abstract}                                                               
\section{Introduction}\label{INTRO}
The fluctuation power spectrum, $P(k)$, of the comoving density
contrast, $\delta(\vec{r})$, is a powerful summary statistic to
explore the second-order clustering properties of cosmic
structures. Its direct relation to theoretical quantities makes it an
ideal tool for the discrimination between different scenarios of
cosmic structure formation and cosmological models in
general. However, measurements give the spatial distribution of
`light' and not the fluctuations of the underlying matter field. For
galaxies the connection between mass and the presence of a stellar
system is complicated because nonlinear gravitational, dissipative,
and radiative processes could lead to a nonlinear biasing up to rather
large scales (e.g., Bertschinger et al. 1997 and references given
therein). For rich clusters the relation between mass and the presence
of such systems is expected to be governed by comparatively simple
biasing schemes (e.g., Kaiser 1984, Bardeen et al. 1986, Mo \& White
1996), mainly driven by gravitation, and only slightly modified by
dissipative processes. In this sense rich clusters of galaxies are
much easier to model and thus `better' tracers of the large-scale
distribution of matter.

Power spectra obtained from optically selected cluster surveys
(Peacock \& West 1992; Einasto et al. 1993; Jing \& Valdarnini 1993;
Einasto et al. 1997; Retzlaff et al. 1998; Tadros, Efstathiou \&
Dalton 1998) are found to have slopes of about $-1.8$ for
$k>0.05\,h\,{\rm Mpc}^{-1}$ and a turnover or some indications for a
turnover at $k\approx 0.03-0.05\,h\,{\rm Mpc}^{-1}$. Contrary to this,
Miller \& Batuski (2000) find no indication of a turnover in the
distribution of Abell richness $\ge 1$ clusters for $k\ge
0.009\,h\,{\rm Mpc}^{-1}$.\footnote{The Hubble constant $H_0$ is given
in units of $h=H_0/(100\,{\rm km}\,{\rm s}^{-1}\,{\rm Mpc}^{-1})$ and
the X-ray source properties (luminosities, etc.) for $h=0.5$, the
cosmic density parameter is $\Omega_0=1$, and the normalized
cosmological constant $\Omega_\Lambda=0$.}  Measurements on scales
$>500\,h^{-1}\,{\rm Mpc}$ or $k<0.013\,h\,{\rm Mpc}^{-1}$ where the
cluster fluctuation signal is expected to be smaller than 1 percent
are, however, extremely sensitive to errors in the sample
selection. The resulting artificial fluctuations increase the measured
power spectral densities and thus prevent any detection of a
decreasing $P(k)$ on these large scales.

The current situation regarding the detection and the location of a
turnover in the cluster power spectra appears to be very controversal
with partially contradicting results.  Physically, the scale of the
expected turnover is closely linked to the horizon scale at
matter-radiation equality. This introduces a specific scale into an
otherwise almost scale-invariant primordial power spectrum and thus
helps to discriminate between the different scenarios of cosmic
structure formation discussed today. The narrow peak found for
Abell/ACO clusters by Einasto et al. (1997) and Retzlaff et al. (1998)
suggests a periodicity in the cluster distribution on scales of
$120\,h^{-1}\,{\rm Mpc}$ and, if representative for the whole cluster
population, is very difficult to reconcile with current structure
formation models. The undoubted identification of the location and
shape of this important spectral feature must, however, include a
clear documentation of the quality of the sample from which it was
derived.

Although the quality of optically selected large-area cluster samples
has been improved during the past years by the introduction of, e.g.,
automatic cluster searches (e.g., Dalton et al. 1992, Lumsden et
al. 1992, Collins et al. 1995) a major step towards precise
fluctuation measurements on very large scales is offered by the use of
X-ray selected cluster samples where also poor systems can be reliably
identified and characterized within the global network of filaments or
other large-scale structures. This is due to several facts.

First, the relation between X-ray luminosity and total cluster mass as
observed (see eq.\,\ref{MLR500}, Reiprich \& B\"ohringer 1999, Borgani
\& Guzzo 2000) and as indicated to first order from the modeling of
clusters as a homologous group of objects scaling with mass (Kaiser
1986), convincingly demonstrates the possibility to select clusters
basically by their mass, although the $1\sigma$ scatter for the
determination of the gravitational mass from X-ray luminosity is still
quite large (about 50 percent).  This is clearly preferable compared
to a selection of clusters by their optical richness, as indicated for
example by the results obtained within the ENACS (Katgert et al. 1996)
where about 10 percent of the Abell, Corwin \& Olowin (1989) clusters
with $z\le 0.1$ (located in the southern hemisphere) do not show any
significant concentration along the redshift direction and must thus
be regarded as spurious.

Second, although the spatial galaxy number density profiles are more
concentrated towards the cluster centres compared to the gas density
profiles, it is the much more centrally peaked X-ray emissivity
profile ($\sim \rho_{\rm gas}^2$) which increases the contrast to the
background distribution and enhances the angular resolution of an
X-ray cluster survey. This decreases the probability of `projection
effects' known to contaminate, e.g., the optically selected Abell/ACO
cluster sample (Lucey 1983, Sutherland 1988, Dekel et al. 1989).

Third, the large-scale variation of galactic extinction modifies the
local sensitivity of cluster detection (for the optical passband see
Nichol \& Connolly 1996). In addition to galactic obscuration galaxies
can be confused with faint stars which reduces the contrast of a
cluster above the background so that the system appears less rich
(Postman, Geller \& Huchra 1986). The resulting artificial distortions
must be reduced because they easily dominate any measured fluctuation
on large scales (e.g., Vogeley 1998). In the following it will be
shown that in X-rays the local survey sensitivity can be readily
computed using the local exposure time of the X-ray satellite and the
local column density of neutral galactic hydrogen, $N_{\rm HI}$.

First results of a power spectrum analysis using X-ray selected
subsamples of the 291 clusters of the ROSAT Bright Survey (Schwope et
al. 2000) are presented in Retzlaff (1999) and Retzlaff \& Hasinger
(2000). For the count rate-limited subsample indications for a
turnover of $P(k)$ at $k=0.05\,h^{-1}\,{\rm Mpc}$ are found. For the
volume-limited subsample the statistical significance of this specific
feature is very weak or almost absent.

In this paper we present the results of a power spectrum analysis
obtained with a sample of 452 ROSAT ESO Flux-Limited (REFLEX) clusters
of galaxies.  A related study of the large-scale distribution of
REFLEX clusters using the spatial two-point correlation function can
be found in Collins et al. (2000). Sect.\,\ref{SAMPLE} gives a brief
overview of the selection of the cluster sample. Sect.\,\ref{SEARCH}
concentrates on the discussion of the overall completeness of the
REFLEX sample, drawing special attention to those selection effects
which might limit the fluctuation measurements on large scales. In
Sect.\,\ref{METHOD} standard methods of power spectral analyses are
applied to estimate $P(k)$. The systematic and random errors are
computed using a set of N-body simulations of an open Cold Dark Matter
(OCDM) model which is shown to give a good though not optimal
representation of the REFLEX sample (Sect.\,\ref{SIMUL}). The results
are shown in Sect.\,\ref{OBSPK} and compared with optically selected
cluster and galaxy samples. In Sect.\,\ref{MODELS} a semi-analytic
model is derived and compared with the observed power spectra of flux-
and of volume-limited subsamples. Sect.\,\ref{DISCUSS} summarizes and
discusses the main results.

\begin{figure*}
\vspace{-5.0cm}
\centerline{\hspace{1.15cm}
\psfig{figure=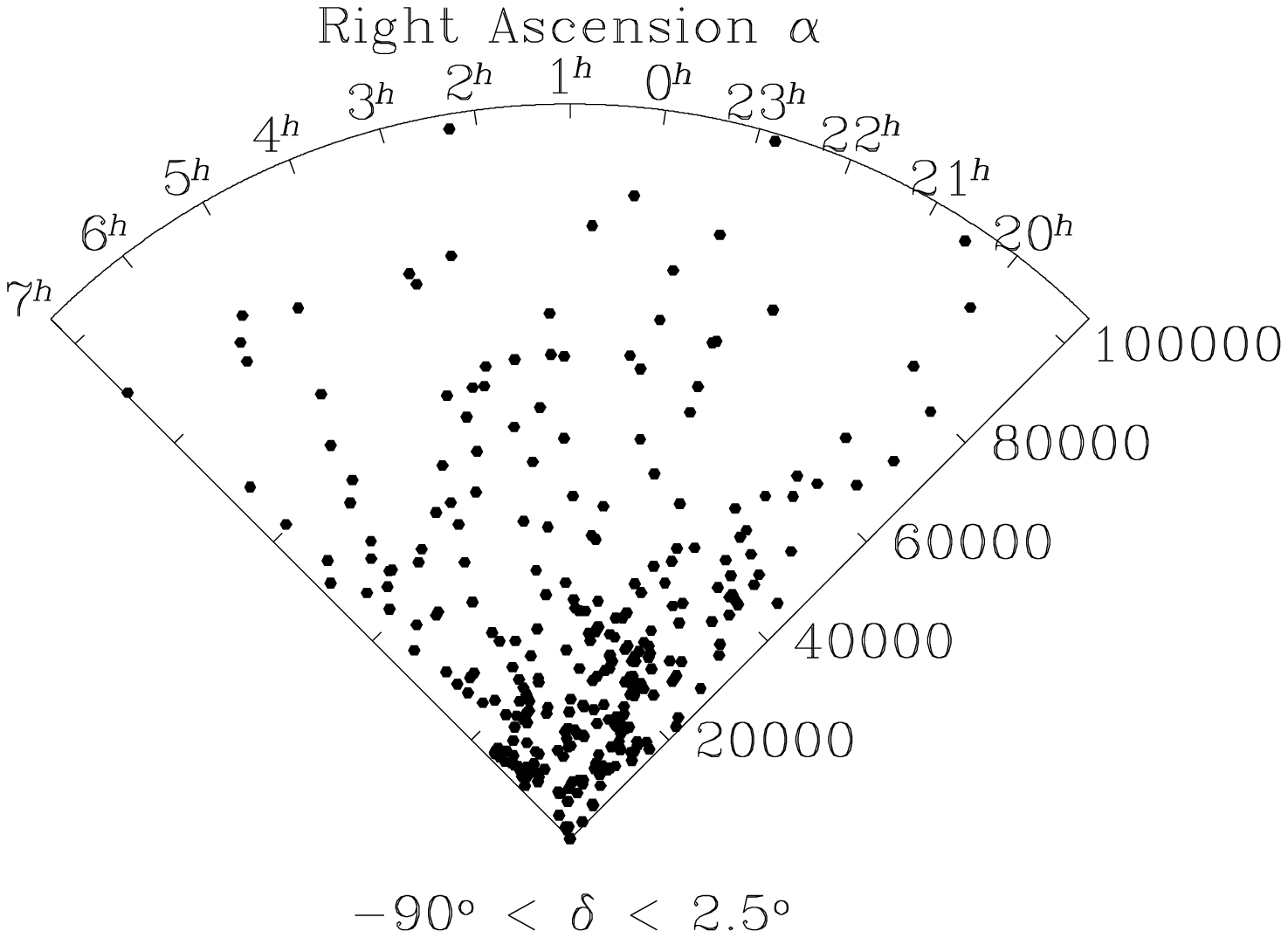,height=12.5cm,width=12.5cm}
\hspace{-3.8cm}
\psfig{figure=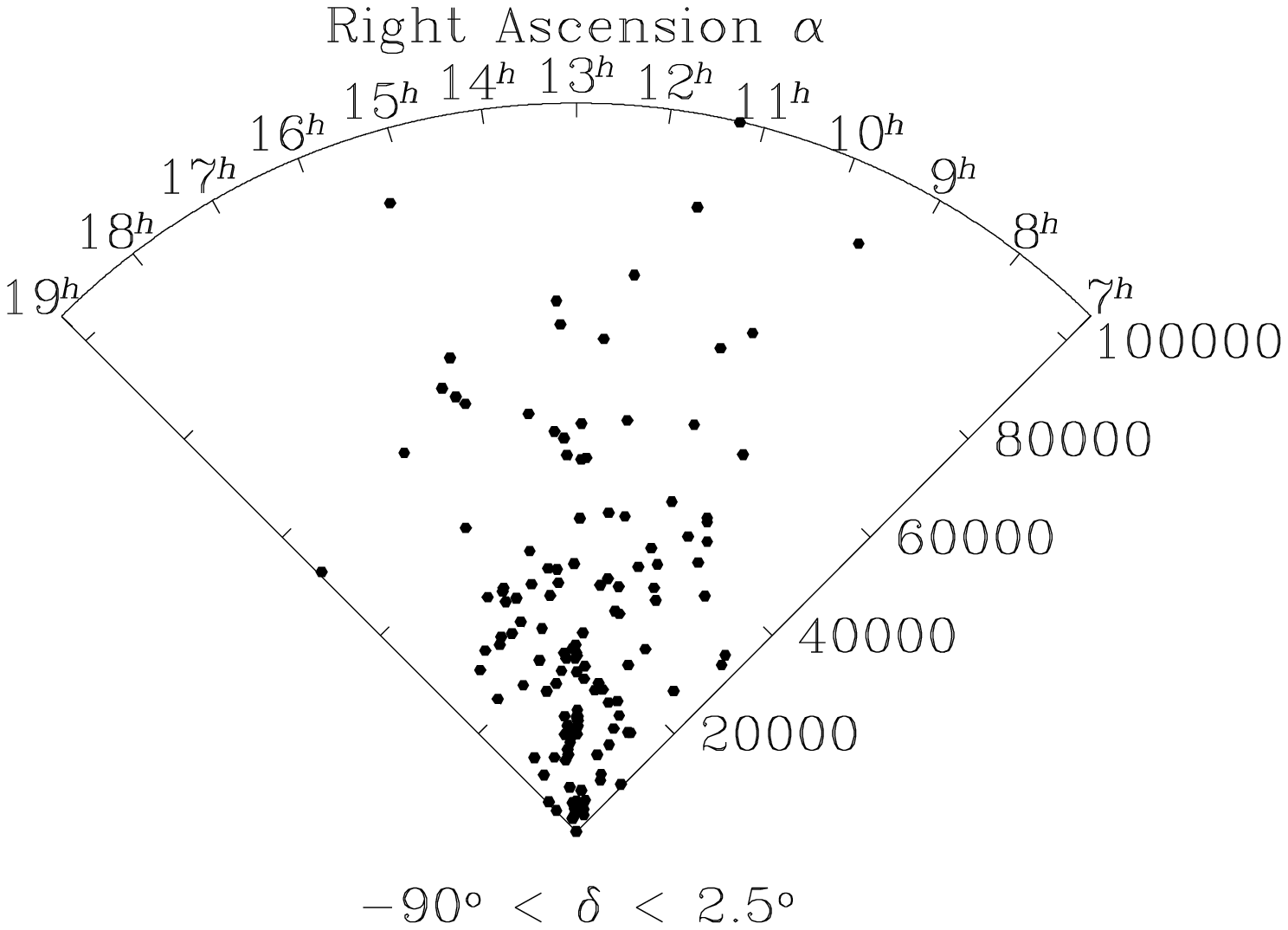,height=12.5cm,width=12.5cm}}
\vspace{-0.5cm}
\caption{\small Spatial distribution of the REFLEX clusters of
galaxies.  Radial axes are given in units of [${\rm km}\,{\rm
s}^{-1}$]. Plotted are all clusters with $z\le 0.35$. Galactic
extinction partially obscures the regions $6^{\rm h}-10^{\rm h}$ and
$16^{\rm h}-20^{\rm h}$. Note the Shapley concentration at ${\rm
R.A.}=13^h$ and $cz\approx 14\,000\,{\rm km}\,{\rm s}^{-1}$ surrounded
by a filament of galaxy clusters.}
\label{FIG_CONE}
\end{figure*}

\section{The REFLEX cluster sample}\label{SAMPLE}
In the following a brief overview of the sample construction is
given. A detailed description of the various reduction steps, the
resulting sample sizes, the methods for the X-ray flux, $S$, and
luminosity, $L_{\rm X}$, computations, the determination of
temperature- and redshift-dependent flux corrections, as well as the
correlation with optical galaxy catalogues, and the computation of the
local survey flux limits (survey sensitivity) can be found in
B\"ohringer et al. (2000a,b).

The REFLEX clusters are detected in the ROSAT All-Sky Survey
(Tr\"umper 1993, Voges et al. 1999). They are distributed over an area
of 4.24\,sr ($13\,924\,{\rm deg}^2$) in the southern hemisphere below
$+2.5$\,deg Declination. To reduce incompleteness caused by galactic
obscuration and crowded stellar fields the sample excludes the area
$\pm 20$\,deg around the galactic plane and 0.0987\,sr at the Small
and the Large Magellanic Clouds, basically following the boundaries of
the corresponding UK Schmidt plates (e.g., Heydon-Dumbleton, Collins
\& MacGillivray 1989).

The sample is based on an MPE internal source catalogue extracted with
a detection likelihood $\ge 7$ from the ROSAT All-Sky Survey
(RASS\,II). 54\,076 southern sources have been re-analysed with the
growth curve analysis method (B\"ohringer et al. 2000b) which is
especially suited to the processing of extended sources. Although the
data were analysed in all three ROSAT energy bands most weight is
given to the hard band (0.5-2.0\,keV) where 60 to 100 percent of the
cluster emission is detected, the soft X-ray background is reduced by
a factor of approximately 4, and the contamination through the
majority of RASS\,II sources is lowest, so that the signal-to-noise
for the detection of clusters is highest.  As expected the new count
rates are systematically higher (up to an order of magnitude) compared
to the count rates given by the standard ROSAT analysis software which
is optimized for the processing of point-like sources.

The low source counts of many RASS sources as well as the limited
spectral resolution of the PSPC do not give enough information for a
proper identification of the sources based only on the X-ray
properties so that additional reduction steps are necessary.  Optical
cluster counterparts are found using counts of COSMOS galaxies
(Heydon-Dumbleton et al. 1989) in concentric rings with different
apertures centered around the X-ray source positions. The probability
thresholds used for the different rings are set low to select also
weak excesses of galaxy surface number densities above background,
introducing a formal sample incompleteness of less than 10 percent.

\begin{figure}
\vspace{0.0cm}
\centerline{\hspace{0.0cm}
\psfig{figure=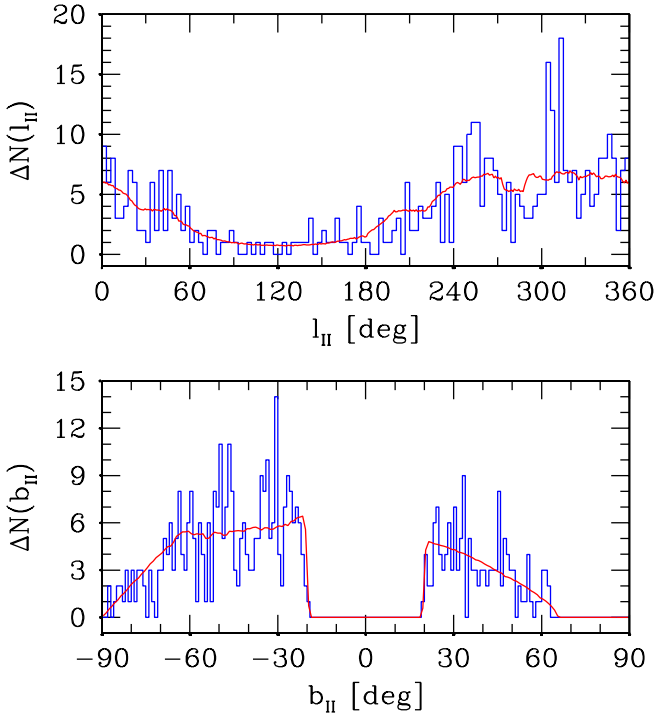,height=10.0cm,width=10.0cm}}
\vspace{0.0cm}
\caption{\small Number of REFLEX clusters of galaxies as a function of
galactic longitude $l_{\rm II}$ and latitude $b_{\rm II}$ (steps)
compared with the number of clusters expected for a random realisation
of the REFLEX selection function (area and sensitivity), computed for
the nominal flux limit $3\times 10^{-12}\,{\rm erg}\,{\rm
s}^{-1}\,{\rm cm}^{-2}$ and the minimum of 10 source counts
(continuous lines). Narrow count bins are chosen to show the effects
of large-scale clustering.}
\label{FIG_LBN}
\end{figure}

The cluster candidates are screened using the X-ray, optical, and
literature data. Obvious multiple detections, and candidates with a
strong point-like contamination (e.g., active galactic nuclei AGN) of
the X-ray flux where the residual flux from the cluster is estimated
to be smaller than the nominal REFLEX flux limit, are removed.  Double
sources are deblended, and count rates measured in the hard band are
converted to {\it unabsorbed} fluxes in the ROSAT band
$(0.1-2.4)$\,keV using standard radiation codes for a thermal spectrum
with temperature $k_{\rm B}T=5.0$\,keV, redshift $z=0$, metal
abundance 0.3 solar units, and local $N_{\rm HI}$ (Dickey \& Lockman
1990, Stark et al. 1992). The internal errors of the measured fluxes
range between 10 and 20 percent.  The effects of a possible
systematic underestimation of the {\it total} fluxes, mainly caused by
the incomplete sampling of the outer parts of the cluster X-ray
emission, are presently investigated (H. B\"ohringer et al., in
preparation). For the present investigation the measured fluxes (not
the total fluxes) are used.

A complete identification of all cluster candidates and a measure of
their redshifts has been performed in the framework of an ESO Key
Programme (B\"ohringer et al. 1998, Guzzo et al. 1999). During this
campaign, 431 X-ray targets were observed with an average of about 5
spectra per target.

The iterative computation of the X-ray luminosity uses in the first
step the redshift and the unabsorbed X-ray flux to give a first
estimate of $L_{\rm X}$. This luminosity and the
luminosity-temperature relation of Markevitch (1998, without
correction for cooling flows) is used to improve the initial
temperature estimate ($5$\,keV). In the next step the count rate-flux
conversion factor is recomputed including now the effects of $z$. The
cluster restframe luminosity is calculated by taking into account the
equivalent to the cosmic K-correction. The X-ray luminosities are
given for the $(0.1-2.4)$\,keV energy band ($h=0.5$). For this band
and for clusters with redshifts $z\le 0.3$ and the temperature
$T=5$\,keV the K-corrections are less than 12 percent.  Note that
the iterative calculation does not introduce any uncertainty in the
selection function, since each value of $L_{\rm X}$ has a unique
correspondence to the first calculated unabsorbed flux and thus to a
uniquely determined survey volume.

\begin{figure*}
\vspace{-8.4cm}
\centerline{\hspace{-0.7cm}
\psfig{figure=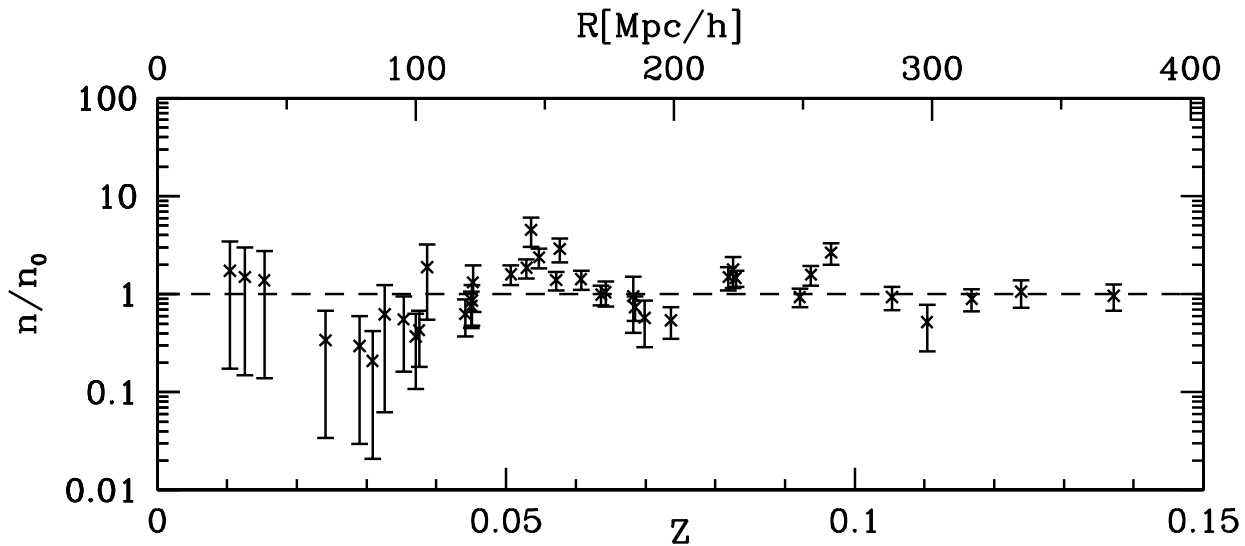,height=19.0cm,width=19.0cm}}
\vspace{-3.2cm}
\caption{\small Normalized comoving cluster number densities as a
function of redshift, $z$, and comoving radial distance, $R$, 
computed with $\Omega_0=1$. Vertical bars represent the formal
$1\sigma$ Poisson errors. Note the quasi-periodic density fluctuations
around an essentially constant mean.}
\label{FIG_CNZ}
\end{figure*}

Adding to the above mentioned selection criteria the nominal flux
limit of the REFLEX sample, $3\times 10^{-12}\,{\rm erg}\,{\rm
s}^{-1}\,{\rm cm}^{-2}$ within the ROSAT energy band $(0.1-2.4)$\,keV,
we find 452 clusters. Of these 449 have measured redshifts, 1 object
is clearly a cluster while 2 are unconfirmed candidates. 65 percent of
the sample are Abell/ACO/Supplement clusters. However, note the
difficulty to compare X-ray flux-limited and richness-limited cluster
samples (see B\"ohringer et al. 2000a for more details). 81 percent of
these clusters show a significant X-ray extent (determined with the
growth curve analysis method). This shows how a selection based solely
on X-ray extent would have missed, given the quality of the RASS\,II
data, a significant percentage of true clusters.  Less than 10 percent
of the REFLEX sources are expected to be significantly contaminated by
unidentified AGN.

Figure\,\ref{FIG_CONE} shows the spatial distribution of the REFLEX
clusters for redshifts $z\le 0.35$.  Galactic extinction partially
obscures the regions $6^{\rm h}-10^{\rm h}$ and $16^{\rm h}-20^{\rm
h}$. The cone diagrams -- although averaged over a large Declination
range -- illustrate the comparatively high sampling rates obtained
with the REFLEX survey. Inhomogeneities in the spatial distribution of
clusters on scales of the order of $100\,h^{-1}\,{\rm Mpc}$ are thus
easily recognized. A detailed analysis of the behaviour of the mean
density and of the topology using Minkowski functionals will be
presented in forthcoming papers. However, the combined effect of the
X-ray flux-limit and the steep X-ray luminosity function (B\"ohringer
et al., in preparation) introduces a systematic dilution of the sample
for larger redshifts. This is an important difference to traditional
optical cluster samples which are up to a certain redshift almost
volume-limited (for given richness).  In the following section the
REFLEX data are tested for artifical number density fluctuations which
could bias fluctuation measurements on large scales.

\section{Tests for artificial density fluctuations}\label{SEARCH}

\subsection{Variation of the local X-ray flux limit across the survey
area (survey sensitivity variations)}\label{VARI}
The local flux limit is determined by the nominal flux limit, the
minimum number of source counts required for a safe detection, the
local exposure time in the RASS\,II, and the local $N_{\rm HI}$
value. According to the resulting {\it survey sensitivity map} for
$\ge 10$ source counts the nominal flux limit $3\times 10^{-12}\,{\rm
erg}\,{\rm s}^{-1}\,{\rm cm}^{-2}$ is reached on 97 percent of the
total survey area. For $\ge 30$ source counts the fraction drops to 78
percent.  For precise fluctuation measurements it is thus necessary to
take into account the local survey sensitivity.

In order to use as many clusters as possible for the fluctuation
measurements all sources with at least 10 source counts in the hard
band are included. Generally, the comparatively low background of the
ROSAT PSPC especially in the hard band allows the detection and the
characterization of sources even with low source counts. In fact, the
number of clusters with 10 to 29 source counts as observed ($N_{\rm
cl}=26$) and as predicted from the subsample of the clusters with at
least 30 source counts ($N_{\rm cl}=37$) suggests a formal
incompleteness of $11\pm 5$ clusters ($1\sigma$ Poisson error, no
cosmic variance) for the subsample with at least 10 source
counts. Assuming that the subsample with at least 30 source counts is
complete this gives a formal overall incompleteness smaller than 3
percent. The corresponding local incompletenesses are expected to be
highest in the areas where the ROSAT satellite passed the radiation
belts in the South Atlantic Anomaly of the Earth's magnetic field.

Random samples are used for the power spectrum analysis giving
Monte-Carlo estimates of the actual REFLEX survey windows
(Sect.\,\ref{METHOD}). They can also be used to test the quality of
the survey selection model. In the following we describe their
construction. The sensitivity map is computed for approximately
$1^\circ\times 1^\circ$ tiles covering the complete sky area $\le
2.5$\,deg Declination. Each of the resulting 21\,529 local selection
functions, $\phi(\vec{r})$, gives the fraction of the X-ray luminosity
function at the comoving distance $\vec{r}$, and thus the number of
expected clusters, $\Delta N(\vec{r})=\bar{n}\phi(\vec{r})\Delta
V(\vec{r})$, down to the local flux limit of the given tile, $S_{\rm
lim}(\alpha,\delta)$, assuming complete randomness. Here, $\bar{n}$ is
the mean comoving cluster number density, $\Delta V(\vec{r})$ the
comoving volume element at $\vec{r}$, and for the given angular
coordinate $(\alpha,\delta)$ of the tile,
\begin{equation}\label{SF}
\phi(\vec{r})\,=\,\frac{\int_{L_{\rm X}(S_{\rm lim}(\alpha,\delta),z)}^\infty\,
\Phi(L_{\rm X})\,dL_{\rm X}}{\int_{L_{\rm X}^{\rm min}}^\infty\,
\Phi(L_{\rm X})\,dL_{\rm X}}\,,
\end{equation}
where $L_{\rm X}^{\rm min}$ is the minimum X-ray luminosity of the
sample.  For the X-ray luminosity function, $\Phi(L_{\rm X})$, we plug
in the empirical estimate of the global REFLEX luminosity function as
determined in B\"ohringer et al. (in preparation). The shape of this
function can be described by a Schechter function with the
characteristic luminosity $L_*=6.04\times 10^{44}\,{\rm erg}\,{\rm
s}^{-1}$ $(h=0.5)$, and the faint-end slope $\alpha=-1.61$ (for  a
minimum of 10 source counts, no deconvolution of measurement error,
see also de\,Grandi et al. 1999). The transformation of the cluster
restframe luminosities into the observer restframe fluxes corresponds
to the reverse of the $S\rightarrow L_{\rm X}$ transformation
described in Sect.\,\ref{SAMPLE}.  This prescription gives a good
representation of the redshift histogram for $L_{\rm X}^{\rm min}=
1.0\times 10^{43}\,{\rm erg}\,{\rm s}^{-1}$ (comparable to the
luminosity of bright Hickson groups).

Figure\,\ref{FIG_LBN} compares the observed cluster surface number
densities as a function of galactic coordinates with the surface
number densities obtained from Monte-Carlo simulations of a {\it
random} distribution of clusters in the REFLEX survey area with at
least 10 X-ray source counts, and modulated by the local variation of
the satellite exposure time and galactic $N_{\rm HI}$ (survey
sensitivity map). The overall agreement is encouraging. The good
statistical coincidence between observed and expected cluster counts
close to the $b_{\rm II}=\pm 20^\circ$ survey boundaries suggests that
the effects of galactic extinction are well represented in the survey
selection model. The remaining local deviations are caused by
large-scale clustering.

\subsection{Variation of the average comoving cluster
number density along the radial direction}\label{SYSTEM}
To test the variation of the average cluster number density along the
radial direction, mean densities are computed for different
volume-limited subsamples taking into account the local survey
sensitivity map by weighting each cluster with the X-ray flux $S$
using the inverse of the fraction of the survey area with a flux limit
below $S$ (effective area). For each subsample the comoving number
densities are normalized to their respective mean density.

Figure\,\ref{FIG_CNZ} shows the normalized comoving number density
computed along the redshift direction for comoving radial distances of
$R\le 400\,h^{-1}\,{\rm Mpc}$ correspond to $z\le 0.15$. Maximum
fluctuations of the order of 3 are found on small scales. They are
successively smoothed out with increasing $R$. The quasi-periodic
density variations have a wavelength of about $150\,h^{-1}\,{\rm
Mpc}$. No related feature is seen in the power spectrum at this scale
(see Sect.\,\ref{OBSPK}). The essentially constant mean comoving
cluster density implies the absence of selection effects
discriminating  against the more distant clusters. Note that the
REFLEX survey covers the southern hemisphere so that a volume with a
radius of $R=400\,h^{-1}\,{\rm Mpc}$ gives a maximum comoving scale
length of about $\lambda=2R=800\,h^{-1}\,{\rm Mpc}$.  Comoving
number densities on Giga parsec scales will be discussed in detail in
B\"ohringer et al. (in preparation).

The huge nearby underdensity centered at $z=0.03$ is also present in
the ESO Slice Project data (Vettolani et al. 1997) as shown in Zucca
et al. (1997) and might be the origin of the observed deficit of
`bright' galaxies in the magnitude number counts as discussed in Guzzo
(1997). The large overdensity region at $z=0.05$ is partially caused
by the Shapley concentration (Fig.\,\ref{FIG_CONE} -- right cone, see
also Scaramella et al. 1989, and Bardelli et al. 1997) and by some
isolated nearby structures located at that distance in the direction
of the South Galactic Pole (Fig.\,\ref{FIG_CONE} -- left cone).

\subsection{Flux-dependent incompletenesses}\label{OVERALL}

Flux-dependent incompletenesses might also lead to systematic errors
in the fluctuation measurements. This section investigates the
presence of this type of incompleteness and its relation to cosmic
variance.

\begin{figure}
\vspace{-1.5cm}
\centerline{\hspace{0.1cm}
\psfig{figure=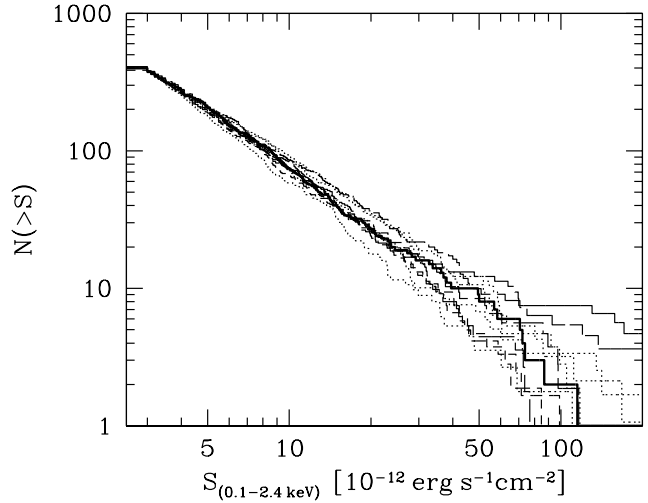,height=9.0cm,width=9.0cm}}
\vspace{-0.5cm}
\caption{\small Cumulative distributions as a function of X-ray flux
for a REFLEX subsample (thick continuous line) and for 10 simulated
OCDM samples (thin continuous, broken, dotted, dashed lines) convolved
with the REFLEX survey sensitivity and normalized to the same number
of clusters. The large scatter of the number counts at high fluxes is
significantly above the formal Poisson expectation and reflects the
effects of cosmic variance.}
\label{FIG_SHISTO}
\end{figure}

For the REFLEX flux range the {\it shape} of the cumulative cluster
number counts as a function of X-ray flux is mainly sensitive to
flux-dependent incompleteness and to the K-correction, weakly
dependent on evolutionary effects, and almost independent of the shape
of a non-evolving X-ray luminosity function (completely independent
for an Euclidean space), the chosen cosmological background model, and
the type of dark matter used in the simulations. The comparison of the
slopes of observed and simulated distributions provides a robust
though model-dependent measure of the relative incompleteness of a
survey (the N-body simulations are described in Sect.\,\ref{NBODY}).

The individual cumulative flux-number counts obtained with 10
statistically independent simulations are shown in
Fig.\,\ref{FIG_SHISTO}.  Cosmic variance modulates the simulated
cluster counts especially for X-ray fluxes $S>5.0\times 10^{-12}\,{\rm
erg}\,{\rm s}^{-1}\,{\rm cm}^{-2}$ yielding slopes between $-1.2$ and
$-1.6$. The fluctuations are caused by the large-scale variations of
comoving cluster number density at small redshifts similar to those
shown in Figs.\,\ref{FIG_CONE} and \ref{FIG_CNZ}. At fainter fluxes
the fluctuations decrease and the slopes of the cumulative
distributions converge to values of about $-1.3$ (note that the
plotted cumulative distributions still contain the effects of the
effective survey area) which is close to the observed slope of $-1.35$
(B\"ohringer et al. 2000a). At this limit the REFLEX sample appears to
be deep and large enough so that the resulting number counts should be
regarded as statistically representative for the local Universe and
not dominated by chance fluctuations. The similarity of observed and
simulated slopes suggests a high overall completeness of REFLEX.

\begin{figure}
\vspace{-1.5cm}
\centerline{\hspace{-1.0cm}
\psfig{figure=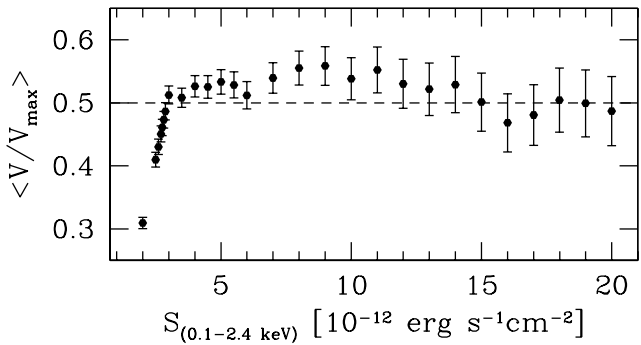,height=10.0cm,width=10.0cm}}
\vspace{-3.5cm}
\caption{\small Mean $V/V_{\rm max}$ values for REFLEX clusters as a
function of X-ray flux.  Error bars are the formal $1\sigma$ Poisson
errors (no cosmic variance). To illustrate the sensitivity of the test
we have computed $V/V_{\rm max}$ also below the nominal flux limit of
$3.0\times 10^{-12}\,{\rm erg}\,{\rm s}^{-1}\,{\rm cm}^{-2}$ where no
REFLEX clusters are present to mimic a simple kind of
incompleteness. Note that the test is performed with all clusters
brighter than a given flux limit $S_{(0.1-2.4\,{\rm keV})}$
introducing a statistical dependency of the averaged $V/V_{\rm max}$
values for different $S$.}
\label{FIG_VVMAX}
\end{figure}

As a second measure of the overall sample incompleteness the $V/V_{\rm
max}$ test (e.g., Schmidt 1968, Avni \& Bahcall 1980) is applied as a
function of the flux limit. Fig.\,\ref{FIG_VVMAX} shows the averaged
$V/V_{\rm max}$ values for different X-ray flux limits. Towards
fainter flux limits the scatter decreases because sample sizes and
volumes increase. At the nominal flux limit, the mean $V/V_{\rm max}$
value is $0.512\pm 0.014$ where the formal error does not include
fluctuations caused by large-scale clustering. We take this
convergence to the ideal case $<V/V_{\rm max}>=0.5$ for a
non-expanding Euclidian universe as a clear sign that at the nominal
flux limit the REFLEX survey volume and sample size is large enough to
cover a representative part of the local Universe with a high sample
completeness and a small sample variance.

To summarize, although it is not the basic aim of the present
investigation to assess the absolute completeness and statistical
representativeness of the REFLEX sample (see B\"ohringer et
al. 2000a), several indications are given that the REFLEX survey is
large enough so that in general the values of statistical quantities
derived from the sample are expected to be not dominated by the effect
of the limited REFLEX survey volume (e.g.,
Figs.\,\ref{FIG_CNZ},\ref{FIG_SHISTO}), and should thus give a useful
characterization of the local Universe. The fluctuation measurements
investigated here will not be dominated by survey incompleteness
(Fig.\,\ref{FIG_VVMAX}) or other artifical large-scale variations out
to radial distances of $400\,h^{-1}\,{\rm Mpc}$ (Fig.\,\ref{FIG_CNZ}).

For the following power spectrum analyses we use different subsamples
which are either flux-limited (abbreviated by F) or volume-limited
(abbreviated by L). Note that the flux limit of the F subsamples is
the nominal flux limit of REFLEX and that most of the F subsamples are
also restricted to different volumes smaller than the total survey
volume (see below). The characteristics of the subsamples are given in
Tab.\,1. The F0 sample contains all clusters with $L_{\rm X}\ge
10^{43}\,{\rm erg}\,{\rm s}^{-1}$, ($h=0.5$), and source counts $\ge
10$. It serves as a reference sample from which the following
subsamples are derived. The subsamples F300 to F800 differ by the
chosen box length, $L$, used for the computation of the Fourier
transforms, varying between $L=300\,h^{-1}\,{\rm Mpc}$ and
$L=800\,h^{-1}\,{\rm Mpc}$. With these subsamples volume-dependent
effects are tested. The volume-limited subsamples L050 and L120 have
luminosity $L_{\rm X}\ge 0.5\times 10^{44}\,{\rm erg}\,{\rm s}^{-1}$
and $L_{\rm X}\ge 1.2\times 10^{44}\,{\rm erg}\,{\rm s}^{-1}$
($h=0.5$), respectively, and are used to analyse the amplitude and
shape of $P(k)$ for clusters with different masses. For the given flux
limit, subsamples with a lower X-ray luminosity cut as used in L050
are surely fluctuation-dominated and can thus not be regarded as
statistically representative. L120 has the largest sample size
attainable for volume-limited REFLEX subsamples. Tab.\,1 gives
comoving cluster number densities and mean cluster-cluster distances
only for the volume-limited subsamples because of the strong dilution
of the flux-limited subsamples and the corresponding large change of
these quantities with increasing redshift.

\begin{table*}
{\bf Tab.\,1.} REFLEX flux-limited (F) and volume-limited (L)
subsamples used for the power spectral analyses of the clusters with
$\ge 10$ source counts and the nominal flux limit $3.0\times
10^{-12}\,{\rm erg}\,{\rm s}^{-1}\,{\rm cm}^{-2}$, and $L_{\rm X}\ge
10^{43}\,{\rm erg}\,{\rm s}^{-1}$. The X-ray luminositities $L_{\rm
X}$ are given for $h=0.5$, box length $L$ used for the Fourier
transformation, averaged comoving cluster number densities $n$, and
mean cluster-cluster distances, $\bar{s}$, in units of $h$. $N_{\rm
CL}$ is the number of clusters in the subsample and $z$ the
redshift. Fluxes and luminosities are given for the ROSAT energy band
$(0.1-2.4)$\,keV.\\
\vspace{-0.5cm}
\begin{center}
\begin{tabular}{lcccccc}
Sample & $L_{\rm X}\ge$ & $z\le$ & $N_{\rm CL}$ & $L$ & $n$ & $\bar{s}$\\
     & $[10^{44}\,{\rm erg}\,{\rm s}^{-1}]$ 
     & & & $[h^{-1}\,{\rm Mpc}]$ & $[h^3\,{\rm Mpc}^{-3}]$
& $[h^{-1}\,{\rm Mpc}]$\\ 
\hline\\
F0 & $0.1$   & $0.460$ & $428$ & -       &- &- \\
F300 & $0.1$ & $0.460$ & $133$ & $300$   &- &- \\
F400 & $0.1$ & $0.460$ & $188$ & $400$   &- &- \\
F500 & $0.1$ & $0.460$ & $248$ & $500$   &- &- \\
F600 & $0.1$ & $0.460$ & $292$ & $600$   &- &- \\
F700 & $0.1$ & $0.460$ & $326$ & $700$   &- &- \\
F800 & $0.1$ & $0.460$ & $341$ & $800$   &- &- \\
L050  & $0.5$ & $0.063$ & $75$  & $400$  & $9.0403\times
10^{-6}$ & $48.0 $ \\
L120  & $1.2$ & $0.093$ & $96$  & $400$  & $3.8312\times
10^{-6}$ & $63.9$ \\ 
\hline
\end{tabular}
\end{center}
\end{table*}
\section{Spectral analyses}\label{METHOD}
\subsection{Formal background}\label{FORMAL}
In the following the spatial distribution of clusters is regarded as a
realisation of a formal point process. The corresponding Fourier
transforms are well-defined in the strict mathematical sense if the
related count measures are approximated by suitably smoothed versions,
allowing the application of the classic Bochner-Khinchin theorem
(e.g., Shiryaev 1995, p.\,287) also for point processes. The
subsequent definition of the classical Bartlett or power spectrum of
point processes via the Fourier transform of reduced second-order
stationary random measures (Ripley 1977), which are closely related to
the two-point (spatial) correlation function, does not cause any
greater difficulties. More details can be found in, e.g., Daley \&
Vere-Jones (1988, Chap.\,11).

\subsection{Spectral estimators}\label{ESTIMATORS}

Problems arise to find unbiased spectral estimators with small
variance and no correlations between power spectral densities obtained
at different wavenumbers $k$. As an example, naive estimators of the
general form (statistical estimates are indicated by the hat symbol)
\begin{equation}\label{SA1}
\hat{P}(k)\,\sim\,\left|\,\frac{1}{N}\,
\sum_{j=1}^{N}\,e^{i\vec{k}\cdot\vec{r}_j}\,-\,W_{\vec{k}}\right|^2\,,
\end{equation}
where $N$ is the number of points which are located at the comoving
positions $\vec{r}_j$, and $W_{\vec{k}}$ the discrete Fourier
transform of the survey window, are basically applied in all
investigations mentioned in Sect.\,\ref{INTRO}. Even for {\it cubic}
survey volumes it leads after the subtraction of the shot noise to the
expectations (abbreviated by the letter E)
\begin{equation}\label{SA2}
\rm{E}\{ \hat{\it P}({\it k}) \}\,=\,\int\,
F(\vec{k}-\vec{k}')\,P(\vec{k}')\,d^3\vec{k}'\,,
\end{equation}
where $F(\cdot)$ is known (in the one-dimensional case) as the
Fej\'{e}r's or Dirichlet's kernel (Percival \& Walten 1993,
Chap.\,6). The resulting systematic distortions of $\hat{P}(k)$ caused
by the sidelobes of $F$ increase with the dynamic range of $P(k)$ and
for small data volumes. Tapering is one method to reduce this type of
leakage (Blackman \& Tukey 1958, p.\,93) but would increase the
variance of $\hat{P}(k)$ as well. Another method is to estimate the
power spectral densities only for those $k$ values where the Fourier
transforms of $F(\cdot)$ are almost zero, namely at the multiples of
the fundamental mode, $k_0=2\pi/L$, where $L$ is the length of the
Fourier box. In order to increase the signal-to-noise the power
spectral densities are averaged over shells in $k$-space with the
thickness $\Delta k=k_0$, centered on the multiples of the fundamental
mode. In a similar way smoothing of $\hat{P}(k)$ with, e.g., Bartlett,
Parzen, or other standard spectral smoothing windows as described in
reference books on Fast Fourier transform would reduce the variance of
$\hat{P}(k)$. However, as shown in Percival \& Walten (1993), either
type of smoothing is critical because especially the central lobe of
the smoothing window introduces a bias $\sim \gamma^2\frac{\partial^2
P(k)}{\partial k^2}$ which is proportional to the squared bandwidth,
$\gamma^2$ (a reasonable estimate of $\gamma$ is given by the
fundamental mode $k_0$), and to the local curvature of $P(k)$.

The leakage introduced by the survey window increases even further for
{\it asymmetric} survey volumes because in this case a unique
fundamental mode does not exist. For almost symmetric windows the
effects are small and might be corrected using the formulae given in
Peacock \& Nicholson (1991) and Lin et al. (1996). For highly
asymmetric windows the whole concept of plane wave approximation
fails. In this case the deconvolution of the survey window function
becomes unreliable below a certain wavenumber, and the best solution
is to resort to survey- and clustering- specific eigenfunctions as
those provided by the Karhunen-Loeve transform (Vogeley \& Szalay
1996). Moreover, the survey volume under consideration might {\it not
be large enough} to cover a representative part of the Universe so
that the resulting `cosmic variance' adds to the technical effects
described above.

Here, for the determination of the power spectrum, two methods are
compared. The first method uses the estimator (Schuecker et
al. 1996a,b)
\begin{equation}\label{PS1A}
\hat{P}(k)\,=\,\frac{V}{\sum_{\vec{k}'}|W_{\vec{k}'}|^2}\,
    \left<\frac{|\hat{\delta}_{\vec{k}}|^2\,-\,\hat{D}}
                    {1-|W_{\vec{k}}|^2}\right>_{|\vec{k}|}\,,
\end{equation}
where the fluctuation amplitudes are corrected for the effects of the
survey window by
\begin{eqnarray}\label{PSS2}
\hat{\delta}_{\vec{k}}\,=\,
\frac{1}{\sum_i [\hat{\phi}(\vec{r}_i)]^{-1}}\,
\sum_{i=1}^{N}\,[\hat{\phi}(\vec{r}_i)]^{-1}\, 
   e^{i\vec{k}\cdot\vec{r}_i}\nonumber \\
\quad\quad\quad\,\,-\,\,
\frac{1}{\sum_j [\hat{\phi}(\vec{r}_j)]^{-1}}\,
\sum_{j=1}^{M}\,[\hat{\phi}(\vec{r}_j)]^{-1}\, 
   e^{i\vec{k}\cdot\vec{r}_j}
\,.
\end{eqnarray}
The estimator of the discreteness noise is
\begin{equation}\label{PS1B}
\hat{D}\,=\,
  \frac{\sum_{i=1}^N(\hat{\phi}(\vec{r}_i))^{-2}}
 {\left[\sum_{i=1}^N(\hat{\phi}(\vec{r}_i))^{-1}\right]^2}
\,+\,
  \frac{\sum_{j=1}^M(\hat{\phi}(\vec{r}_j))^{-2}}
 {\left[\sum_{j=1}^M(\hat{\phi}(\vec{r}_j))^{-1}\right]^2}\,.
\end{equation}
The squared differences of the discrete Fourier transforms of the
observed (inhomogeneous) and of the random distributions, both
corrected for shot noise, are averaged over different directions and
weighted by $(1-|W_{\vec{k}}|^2)$ reducing the effects of the errors
in the mean number density (Peacock \& Nicholson 1991). The power
spectral densities must be normalized by the volume $V$ used to
compute the Fourier transforms and by the total power of the Fourier
transformed survey window. Whereas the number of observed objects $N$
is fixed by the sample, the number of points used for the random
sample $M$ should be large enough so that their shot noise
contributions can be subtracted with high accuracy. Both the observed
and the random samples have the same position-dependent selection
function, $\phi(\vec{r})$.

The second method to determine the power spectrum averages the
fluctuation power over $N_k$ modes per $k$ shell (Feldman, Kaiser \&
Peacock 1994),
\begin{equation}\label{FKP1}
\hat{P}(k) =
\frac{1}{N_k}\,\sum_{\vec{k}}|\hat{\cal{F}}(\vec{k})|^2-\hat{\cal{D}}\,,
\end{equation}
where the window-corrected Fourier-transformed density contrasts are
given in a similar way as before,
\begin{eqnarray}\label{FKP2}
\hat{\cal{F}}(\vec{k})\,=\,\sum_{i=1}^N\,w(\vec{r}_i)
\,e^{i\vec{k}\cdot\vec{r}_i}\,-\,\alpha\,\sum_{j=1}^M\,w(\vec{r}_j)
\,e^{i\vec{k}\cdot\vec{r}_j}\,.
\end{eqnarray}
The total shot noise is estimated by 
\begin{equation}\label{FKP3}
\hat{\cal{D}}\,=\,\alpha(1+\alpha)\,\sum_{j=1}^{M}w^2(\vec{r}_j)\,
e^{-\vec{k}\cdot\vec{r}_j}\,,
\end{equation}
where $\alpha=N/M$. For Gaussian fluctuations the weights
$w(\vec{r})\,=\,\frac{1}{1\,+\,n(\vec{r})\,P(k)}$ minimize the
variance of the estimator, however, they require the {\it a priori}
knowledge of $P(k)$, that is, the quantity one wants to measure, in
addition to a fair estimate of the mean density,
$n(\vec{r})$. Reasonable results are attainable if $n(\vec{r})$ is
estimated by the observed luminosity function or by smoothed empirical
$z$ histograms (Sect.\,\ref{VARI}) and the sensitivity map of the
survey, and if $P(k)$ is approximated by a constant power spectrum,
$P(k)=P_0={\rm const}$.

\section{Test of the spectral analyses}\label{SIMUL}

\subsection{General tests}\label{GTESTS}

\begin{figure}
\vspace{-0.5cm}
\centerline{\hspace{0.0cm}
\psfig{figure=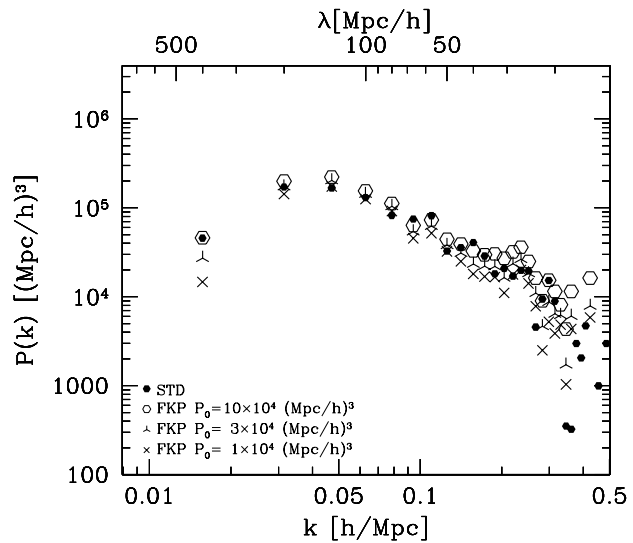,height=8.5cm,width=8.5cm}}
\vspace{-0.5cm}
\caption{\small Power spectral densities obtained with eq.\,\ref{PS1A}
(`standard method' STD) and with eq.\,\ref{FKP1} (`Feldman, Kaiser,
Peacock method' FKP) for a flux-limited REFLEX subsample with $N=188$
clusters within a cubic volume $V=(400\,h^{-1}\,{\rm Mpc})^3$. The
radial parts of the random samples used to estimate the survey window
are computed with smoothed empirical $z$ histograms.}
\label{FIG_FKP}
\end{figure}

The first test concerns the choice of the spectral estimator used for
the analyses of the REFLEX data. Fig.\,\ref{FIG_FKP} compares the
estimates obtained with eqs.\,(\ref{PS1A}) and (\ref{FKP1}). The power
spectral densities are computed for a flux-limited REFLEX subsample in
a cubic box with a length of $L=400\,h^{-1}\,{\rm Mpc}$ using a
standard FFT algorithm on a $128^3$ grid for $N=188$ REFLEX clusters
and for $M=2.0\times 10^6$ random particles. The differences between
the power spectral densities obtained with eqs.\,(\ref{PS1A}) and
(\ref{FKP1}) and the differences between the power spectra obtained
with (\ref{FKP1}) for different $P_0$ are small compared to the errors
introduced by the sample itself (see Sect.\,\ref{NBODY}). We choose
(\ref{PS1A}) for the spectral analyses because the exploration of the
REFLEX data should start with a minimum of pre-assumptions about
$P(k)$. Moreover, the REFLEX survey volume is comparatively symmetric
so that in addition to the window correction term in eq.\,(\ref{PS1A})
no specific deconvolutions are performed. The remaining effects of the
window functions are checked using the results obtained with N-body
simulations (see Sect.\,\ref{NBODY}).

\begin{figure}
\vspace{-0.5cm}
\centerline{\hspace{-0.5cm}
\psfig{figure=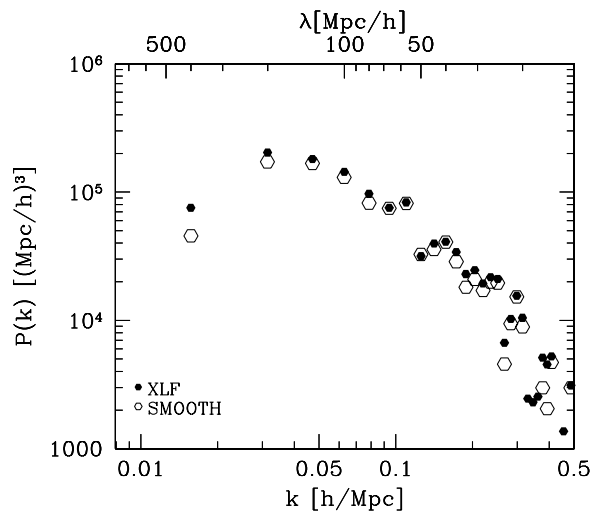,height=8.0cm,width=8.0cm}}
\vspace{-0.6cm}
\caption{\small Fluctuation power spectral densities, $P(k)$, as a
function of comoving wavenumber, $k$, corresponding to the wavelength
$\lambda=2\pi/k$.  Compared are REFLEX $P(k)$ obtained with two
different methods to compute the radial part of the reference random
sample. Filled symbols are obtained with random samples based on the
X-ray luminosity function (XLF), open symbols are based on smoothed
redshift histograms of observed samples (SMOOTH).}
\label{FIG_COMP_XLF_SMOOTH}
\end{figure}

To test the robustness of the method applied for the computation of
the radial parts of the random samples (see Sect.\,\ref{VARI}) the
empirical $z$ histogram is determined for different flux limits and
smoothed with the biweight kernel (corrected for edge effects) using
the standard deviation $\sigma_z=0.03$ to reduce the large-scale
fluctuations. The filtered redshift distributions give an alternative
representation of the radial selection functions (after proper
normalization with the comoving volume elements and the survey
sensitivity map). The local redshift distribution of the random sample
as well as the local radial selection function is then estimated by
the Monte-Carlo method.

\begin{figure}
\vspace{-3.2cm}
\centerline{\hspace{-0.7cm}
\psfig{figure=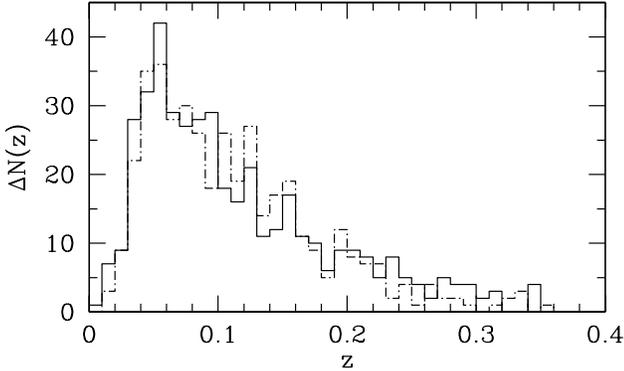,height=9.0cm,width=9.0cm}}
\vspace{-0.8cm}
\caption{\small REFLEX (continuous) and simulated (dashed-dotted) 
cluster redshift histograms.}
\label{FIG_NZ}
\end{figure}

\begin{figure}
\vspace{0.0cm}
\centerline{\hspace{3.0cm}
\psfig{figure=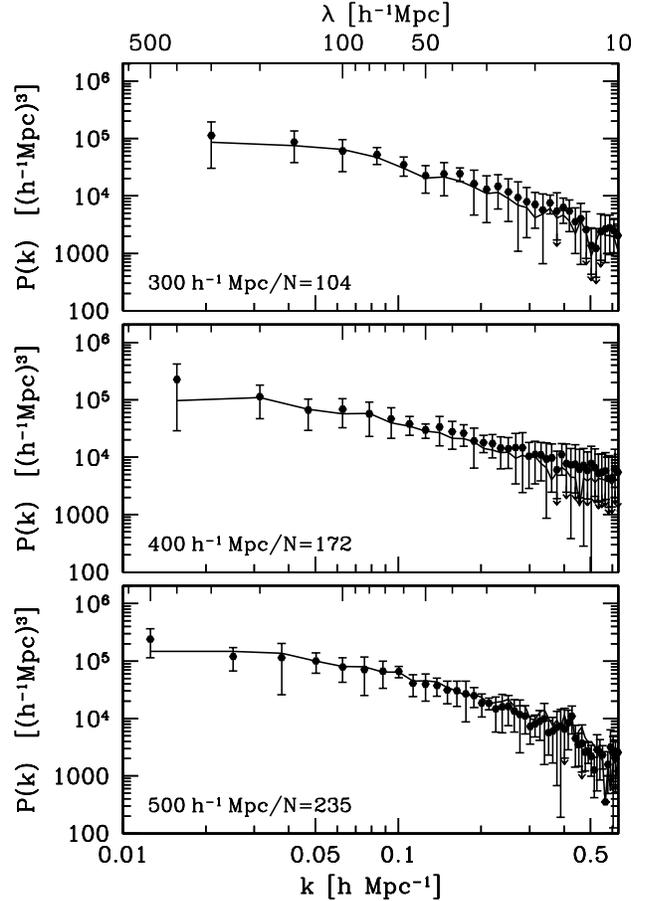,height=12.0cm,width=12.0cm}}
\vspace{0.0cm}
\caption{\small Power spectra for simulated (OCDM) flux-limited
subsamples in different volumes. Filled symbols give the average power
spectral densities obtained by imposing the REFLEX survey conditions,
continuous lines the average power spectral densities obtained for
all-sky cluster surveys with uniform survey sensitivities and no
galactic extinction, but with the same X-ray luminosity function as
the corresponding REFLEX subsamples (error bars omitted). The error
bars are the $1\sigma$ standard deviations of the power spectral
densities obtained with 10 REFLEX-like simulations. Error bars
exceeding the plotted $P(k)$ range are shown by arrows. The size of
the Fourier box and the average number of simulated clusters located
in the box is given in the lower left of each panel.}
\label{FIG_PK_TEST}
\end{figure}

As an example, the power spectral densities shown as open symbols in
Fig.\,\ref{FIG_COMP_XLF_SMOOTH} are computed with random samples based
on smoothed empirical $z$ distributions, the filled symbols with
random samples based on the REFLEX X-ray luminosity function. It is
seen that the power spectral densities obtained with the smoothing
method are systematically smaller up to factors reaching 1.6 at the
largest scales. The differences at small scales are mainly caused by
the poor sampling of density waves by the REFLEX clusters. We regard
the luminosity function method to be more reliable, especially on
large scales (and for small sample sizes): smoothing out all
fluctuations is almost impossible, especially on large scales, so that
the resulting spectra have systematically smaller amplitudes as
illustrated by Fig.\,\ref{FIG_COMP_XLF_SMOOTH}. In the following all
REFLEX power spectra excluding those shown in Fig.\,\ref{FIG_FKP} are
obtained by using the luminosity function to compute the radial part
of the random samples.

\begin{table*}
{\bf Tab.\,2:}  REFLEX power spectral densities obtained for different
REFLEX subsamples, indicated by the index (see Tab.\,1) of the power
spectral densities, $P$. The errors are the formal $1\sigma$ standard
deviations as adapted from 10 OCDM simulations.
\[
\hspace{0.5cm}\begin{array}{|rrr|rrr|rrr|r|rr|rr|}
\hline
\noalign{\smallskip}
k & P_{\rm F300}(k) & \sigma(P) & k & P_{\rm F400}(k) & \sigma(P) &
k & P_{\rm F500}(k) & \sigma(P) & k & P_{\rm L050}(k) & \sigma(P) &
P_{\rm L120}(k) & \sigma(P) \\ 
\noalign{\smallskip}
\hline 
 0.0209  & 225005 & 165087 & 0.0157 &  75312 &  65600 & 0.0126 & 573540 & 301495 & 0.0157 & 208388 & 122391 & 392794 & 301752\\ 
 0.0419  & 295582 & 166814 & 0.0314 & 203050 & 119346 & 0.0251 & 261792 & 114413 & 0.0314 & 283215 & 203881 & 453834 & 346729\\
 0.0628  & 130385 &  73346 & 0.0471 & 180709 & 100744 & 0.0377 & 355915 & 274882 & 0.0471 & 423245 & 227117 & 482384 & 366700\\
 0.0838  &  75460 &  24608 & 0.0628 & 143612 &  75696 & 0.0503 & 156029 &  59838 & 0.0628 & 218190 &  82455 & 428681 & 230770\\
 0.1047  &  71151 &  26249 & 0.0785 &  96808 &  57489 & 0.0628 &  50935 &  23333 & 0.0785 & 187649 &  97309 & 355547 & 293808\\
 0.1257  &  54969 &  27508 & 0.0942 &  75444 &  41117 & 0.0754 & 123809 &  80121 & 0.0942 & 105025 &  68010 & 281954 & 268664\\
 0.1466  &  54289 &  31769 & 0.1100 &  83073 &  30014 & 0.0880 &  82749 &  40760 & 0.1100 & 101841 &  65454 & 223302 & 172965\\
 0.1676  &  47469 &  12606 & 0.1257 &  31639 &   8920 & 0.1005 &  82533 &  18233 & 0.1257 &  70705 &  30683 & 131183 &  95468\\
 0.1885  &  34690 &  24804 & 0.1414 &  39688 &  21014 & 0.1131 &  48720 &  20268 & 0.1414 &  63820 &  36793 & 100848 &  62951\\
 0.2094  &  33145 &  24350 & 0.1571 &  40994 &  20666 & 0.1257 &  35619 &  17539 & 0.1571 &  51805 &  30448 & 121818 &  90020\\
 0.2304  &  33044 &  19821 & 0.1728 &  34105 &  13942 & 0.1382 &  68543 &  23563 & 0.1728 &  62935 &  60617 & 110103 &  60398\\
 0.2513  &  23607 &  16395 & 0.1885 &  22950 &  15291 & 0.1508 &  45761 &  19363 & 0.1885 &  51164 &  34999 &  93618 &  77630\\
 0.2723  &  15823 &  13983 & 0.2042 &  24635 &   8116 & 0.1634 &  30936 &  14519 & 0.2042 &  47937 &  25678 &  70140 &  28187\\
 0.2932  &  16462 &  12553 & 0.2199 &  19413 &   8452 & 0.1759 &  48897 &  32916 & 0.2199 &  55192 &  41352 &  84617 &  38806\\
 0.3142  &  19420 &  11962 & 0.2356 &  21710 &  12222 & 0.1885 &  20998 &   7960 & 0.2356 &  48857 &  33925 &  81083 &  70289\\
 0.3351  &  19051 &  16859 & 0.2513 &  21041 &  11528 & 0.2011 &  26694 &  11482 & 0.2513 &  41602 &  20935 &  75968 &  79354\\
 0.3560  &   6133 &   2189 & 0.2670 &   6694 &   5141 & 0.2136 &  10775 &   2504 & 0.2670 &  38002 &  21234 &  65442 &  51403\\
 0.3770  &  11933 &  13125 & 0.2827 &  10279 &   8592 & 0.2262 &   7637 &   4614 & 0.2827 &  42094 &  28279 &  81393 &  68258\\
 0.3979  &   7290 &   3426 & 0.2985 &  15554 &  11277 & 0.2388 &  10707 &   6913 & 0.2985 &  36277 &  17070 &  70096 &  53046\\
         &        &        & 0.3142 &  10472 &   7294 & 0.2513 &  10938 &   5870 & 0.3142 &  35870 &  19565 &  61827 &  60854\\
         &        &        &        &        &        & 0.2639 &   9894 &   5652 & 0.3300 &  33286 &  17785 &  49837 &  18589\\
         &        &        &        &        &        & 0.2765 &  10248 &   8046 & 0.3456 &  20573 &  16773 &  32071 &  40253\\  
         &        &        &        &        &        & 0.2890 &   8869 &   4518 & 0.3613 &  21151 &  15062 &  34525 &  25173\\
         &        &        &        &        &        & 0.3267 &   8259 &   4428 & 0.3770 &  19253 &  19786 &  43460 &  30951\\
         &        &        &        &        &        &        &        &        & 0.3927 &  20295 &  20823 &  29257 &  21929\\
\noalign{\smallskip}
\hline
\end{array}
\]
\label{PKDATA}
\end{table*}

\subsection{N-body simulations}\label{NBODY}

Systematic and random errors of $\hat{P}(k)$ are investigated using a
set of statistically independent cluster distributions obtained from
realistic N-body simulations, transformed into redshift space, and
modified according to the REFLEX survey selection as summarized by the
survey sensitivity map. In the following a brief overview of some
technical aspects of the simulations are given. A more detailed
description will be presented in the second paper on the REFLEX power
spectrum.

The simulations are performed using a standard PM code (Hockney \&
Eastwood 1988) with $256^3$ particles in a $(500\,h^{-1}\,{\rm
Mpc})^3$ box on a $512^3$ grid giving the force resolution $\Delta x
\approx 1\,h^{-1}\,{\rm Mpc}$. Ten OCDM models are simulated with the
parameters $h=0.60$, cosmic density parameter of matter,
$\Omega_0=0.40$, cosmological constant $\Omega_\Lambda=0$, cosmic
density parameter of baryons, $\Omega_{\rm b}=0.05$ (this corresponds
to an estimate of Burbles \& Tytler 1998) and $\sigma_8=0.80$. The
transfer function was calculated with the Boltzmann code CMBFAST of
Seljak \& Zaldamiaga (1996). The normalization is so as to provide the
correct cluster abundance satisfying both the relation given in Eke,
Cole \& Frenk (1996) and in Viana \& Liddle (1996). We chose this
model because it gives a good representation of the REFLEX data and
thus realistic error estimates. However, any other model with a
similar power spectrum could do the job as well. The mass resolution
is $8.4\times 10^{11}\,h^{-1}\,M_\odot$. Each simulation starts at the
redshift $z=50$ (initial perturbations imposed on the `glass-like'
initial load using the Zel'dovich approximation) and ends after 245
time steps (increment of the scale factor $\Delta a=0.004$). Several
replicants of the same simulation are combined using periodic boundary
conditions to compare results on larger scales. However, only for
scales $\le 500\,h^{-1}\,{\rm Mpc}$ statistically independent
measurements can be obtained.

For the identification of the clusters the friend-of-friend method
(Davis et al. 1985) is used with the linking parameter $b=0.16$ to
pick up virialized structures. The total cluster masses are computed
within the radius where the average density ratio is
$\bar{\rho}(r_{500})/\rho_{\rm critical}=500$ using only those
clusters with at least 10 particles. 
The masses are transformed into luminosities with the empirical mass
-- X-ray luminosity relation for $r_{500}$ from (Reiprich \&
B\"ohringer 1999),
\begin{equation}\label{MLR500}
\frac{M}{h^{-1}\,{\rm M_\odot}}\,=\,2.52\times 10^{14}\,
\left(\frac{L_{\rm x}}{10^{44}\,h^{-2}\,{\rm erg}\,{\rm
s}^{-1}}\right)^{0.81}\,.
\end{equation}
The individual cluster masses used to derive (\ref{MLR500}) show a
$1\sigma$ scatter of about 50 percent.  The $r_{500}$ radius and
the OCDM model parameters mentioned above lead to realistic spatial
cluster distributions, especially redshift histograms
(Fig.\,\ref{FIG_NZ}) and sample sizes, and thus to realistic error
estimates for $\hat{P}(k)$. Whereas in the present investigation the
model parameters taken from the literature are not changed, studies
are in preparation using the `standard' radius $r_{200}$ (see
Sect.\,\ref{SEMOD}) and different types of structure formation models
to adjust the parameter values in order to reconcile the models with
the observations. We apply the above equation assuming no intrinsic
scatter of the mass-luminosity relation. A simulated cluster is
rejected if its flux is below the local flux limit given by the survey
sensitivity map (for 10 source counts). In this way the simulated
cluster sample follows the same sensitivity pattern as the observed
REFLEX sample. Finally, sample sizes are adjusted using the additional
lower luminosity limit $L_{\rm X}=3\times 10^{43}\,{\rm erg}\,{\rm
s}^{-1}$ ($h=0.5$), corresponding to a minimum of 46 particles per
cluster. This approximate modeling gives realistic spatial cluster
distributions and is enough for the error estimation of $\hat{P}(k)$.

As a brief overview, Fig.\,\ref{FIG_NZ} illustrate the similarity of
the observed and simulated cluster samples (see also the cumulative
flux-number counts in Fig.\,\ref{FIG_SHISTO}). The model parameters
are not yet fully optimized to fit the observed data in detail. A more
quantitative comparison is given in Sect.\,\ref{MODELS}.

Figure\,\ref{FIG_PK_TEST} shows the power spectra obtained from (a)
simulated data under realistic REFLEX survey conditions (filled
symbols), and (b) simulated all-sky cluster surveys with uniform
survey sensitivities and no obscuration due to galactic extinction
(continuous lines). The error bars of the latter measurements are
omitted.  The X-ray luminosity functions of the two sets of
simulations are identical so that it is straightforward to test
whether the power spectral estimator gives the correct power spectrum:
after the correct elimination of the effects of the REFLEX survey
window (see eq.\,\ref{PSS2}) the resulting power spectra (shape and
amplitude) of realistic and all-sky simulations should be the
same. The simulations correspond to the F300 to F500 REFLEX subsamples
(Tab.\,1). The errors shown in Fig.\,\ref{FIG_PK_TEST} represent the
$1\sigma$ standard deviations obtained from a set of 10 different OCDM
realizations. For these simulations a maximum of $\hat{P}(k)$ is
expected at $k\approx 0.02\,h\,{\rm Mpc}^{-1}$. Note that the shape
and amplitude of the ideal power spectrum can be recovered under
REFLEX conditions in all volumes analyzed. Some extra power is seen at
the fundamental mode in the $400$ and $500\,h^{-1}\,{\rm Mpc}$
results, however within the $1\sigma$ range. Note that the
$(500\,h^{-1}\,{\rm Mpc})^3$ simulations do not give a good
representation of the fundamental mode at $k=2\pi/500\,h\,{\rm
Mpc}^{-1}$ -- only 3 modes are realized per simulation -- and do not
include any fluctuations on larger scales, so that one should take the
error bars obtained at the simulation limit with
caution. Nevertheless, the overall agreement of the power spectra
obtained under REFLEX and ideal survey conditions suggests that no
significant systematic errors of $\hat{P}(k)$ are expected.
\section{Observed power spectra}\label{OBSPK}
\subsection{Exploring the general shape of $\hat{P}(k)$}\label{SHAPE}
\begin{figure}
\vspace{-1.5cm}
\centerline{\hspace{0.0cm}
\psfig{figure=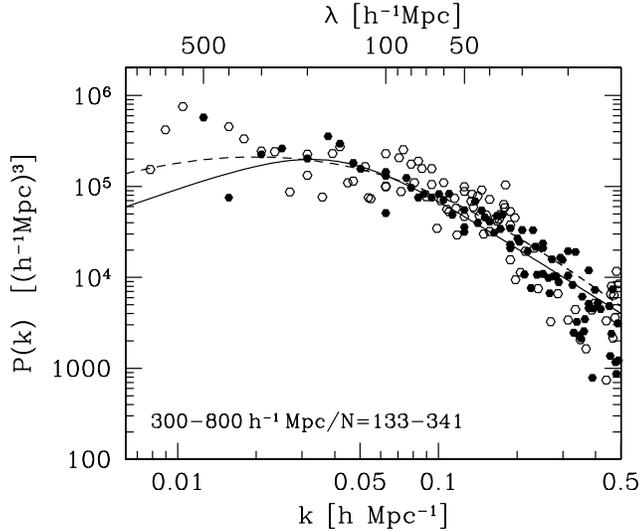,height=9.0cm,width=9.0cm}}
\vspace{-0.8cm}
\caption{\small REFLEX power spectra of the flux-limited subsamples
F300 to F500 (filled symbols, monitored by the N-body simulations) and
F600 to F800 (open symbols) in volumes with box lengths between $300$
and $800\,h^{-1}\,{\rm Mpc}$ (no corrections for differences in
effective biasing). For reference, the spectral fits obtained with the
phenomenological model (continuous line) and with the CDM-like model
(dashed line) using the subsamples F300 to F800 are superposed.}
\label{FIG_PSA_HUGE}
\end{figure}
\begin{figure}
\vspace{0.0cm}
\centerline{\hspace{3.0cm}
\psfig{figure=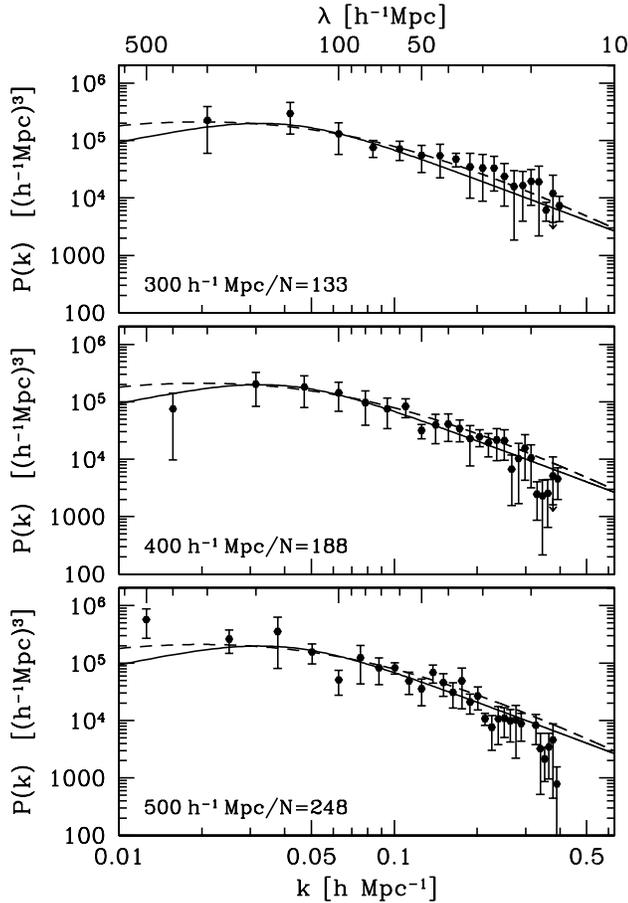,height=12.0cm,width=12.0cm}}
\vspace{0.0cm}
\caption{\small REFLEX power spectra of the flux-limited subsamples
F300 to F500. The box lengths and the number of clusters used for the
power spectrum estimation are given in each panel in the lower
left. The bars represent the $1\sigma$ errors adapted from N-body
simulations. The fits of a linear power spectrum model (dashed lines)
and of the phenomenological model (continuous line) using the
subsamples F300 to F800 are superposed.}
\label{FIG_PSA_ALL}
\end{figure}
\begin{figure}
\vspace{0.0cm}
\centerline{\hspace{3.0cm}
\psfig{figure=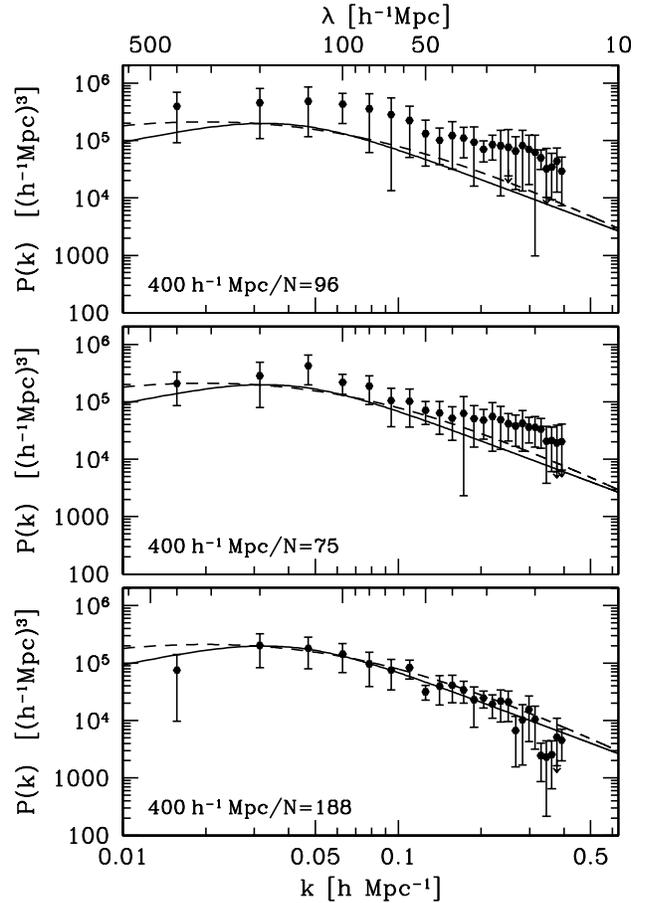,height=12.0cm,width=12.0cm}}
\vspace{0.0cm}
\caption{\small REFLEX power spectra of the volume-limited subsamples
L120 (upper panel), and L050 (middle panel), and of the flux-limited
subsample F400 (lower panel). L120 contains clusters with a brighter
lower X-ray luminosity compared to L050 (and F400).  The bars
represent the $1\sigma$ errors adapted from N-body simulations. For
reference the fits obtained with the phenomenological model
(continuous lines) and with the CDM-like model (dashed lines) using
the subsamples F300 to F800 are superposed. The amplitudes of $P(k)$
increase with increasing lower luminosity limit as expected by
standard biasing schemes.}
\label{FIG_PSA_VOL}
\end{figure}
\begin{figure*}
\vspace{-5.0cm}
\centerline{\hspace{-2.0cm}
\psfig{figure=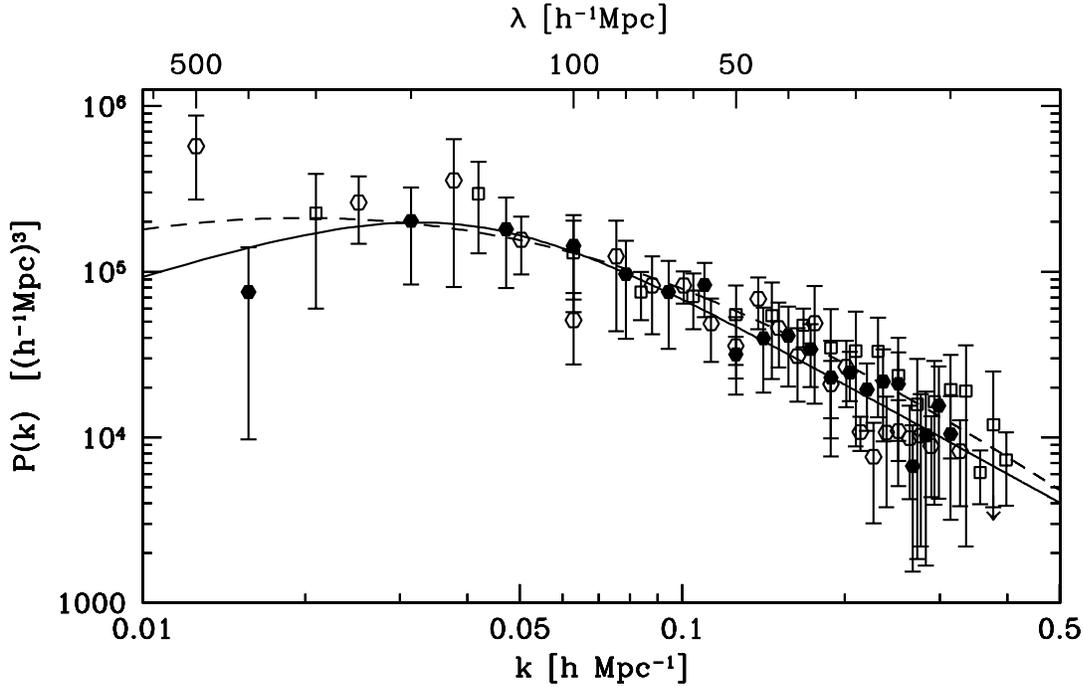,height=16.0cm,width=16.0cm}}
\vspace{-1.9cm}
\caption{\small Combined REFLEX power spectrum obtained with the
flux-limited subsamples F300 (open squares), F400 (filled hexagons),
F500 (open hexagons), and their standard $1\sigma$ deviations adapted
from N-body simulations. Not shown are the power spectral densities
with wavenumbers $k\ge 0.4\,h\,{\rm Mpc}^{-1}$ because the
corresponding density waves are only sparsely sampled by REFLEX. The
continuous line is a fit of the phenomenological model, the dashed
line the CDM-like model fit using the flux-limited subsamples F300 to
F800. The power spectral density at $500\,h^{-1}\,{\rm Mpc}$ is part
of the extra power detected at $k\approx 0.01\,h\,{\rm Mpc}^{-1}$ and
might already be biased.}
\label{FIG_POWER}
\end{figure*}

Many variants of cosmic structure formation models discussed today
predict an almost linear slope of the power spectrum on scales
$<40\,h^{-1}\,{\rm Mpc}$ and a turnover into the primordial regime
between $100$ and $300\,h^{-1}\,{\rm Mpc}$. To summarize our
measurements in this interesting scale range, Fig.\,\ref{FIG_PSA_HUGE}
shows the power spectral densities obtained with the {\it
flux-limited} REFLEX subsamples F300 to F800. The volumes differ by a
factor 19, enabling tests of possible volume-dependent effects
(Sect.\,\ref{CHAR}). The superposed continuous and dashed lines in
this and the following figures of this section are always the
same. Their computation and interpretation is described in
Sect.\,\ref{CHAR}. In the following they may serve as a mere reference
to compare the power spectra obtained with the different REFLEX
subsamples listed in Tab.\,1. Fig.\,\ref{FIG_PSA_ALL} gives a more
detailed view of the spectra obtained with the subsamples F300 to F500
in volumes which are monitored by our N-body
simulations. Fig.\,\ref{FIG_PSA_VOL} compares the spectra obtained for
the {\it volume-limited} subsamples L050 and L120 with the spectrum
obtained for the {\it flux-limited} subsample F400, all spectra are
estimated within the same Fourier volume. Finally,
Fig.\,\ref{FIG_POWER} shows the combined power spectrum obtained with
the subsamples F300 to F500 which we regard as the basic result of the
REFLEX power spectrum analyses. The values of the power spectral
densities obtained with the subsamples F300-F500, L050, and L120 with
the errors estimated with the N-body simulations are given in
Tab.\,2. In the following a few more detailed remarks are given.

Figure\,\ref{FIG_PSA_HUGE} shows the superposition of the power
spectra obtained with the flux-limited subsamples F300 to F800 in the
comoving volumes ranging from $(300\,h^{-1}\,{\rm Mpc})^3$ to
$(800\,h^{-1}\,{\rm Mpc})^3$. The data are not corrected for
sample-to-sample variations of the effective biasing (see
Sect.\,\ref{MODELS}) so that the effective variance among the
estimates is possibly smaller than that displayed by the figure. The
point distribution outlines a corridor which can be  separated
into three parts. For $k>0.1\,h\,{\rm Mpc}^{-1}$ the power spectral
densities decrease approximately as $k^{-2}$. Between $0.02\le k\le
0.1\,h\,{\rm Mpc}^{-1}$ the spectra bend into a flat distribution. The
N-body simulations give $1\sigma$ standard deviations between 30 and
80 percent (including cosmic variance) in this scale range as shown in
Figs.\,\ref{FIG_PSA_ALL}, \ref{FIG_PSA_VOL}, \ref{FIG_POWER}. For
$k<0.02\,h\,{\rm Mpc}^{-1}$ a second maximum is seen at $k\approx
0.01\,h\,{\rm Mpc}^{-1}$. We did not perform N-body simulations for
such large scales. However, the delete-d jackknife resampling method
(a variant of the boostrap method where the creation of artifical
point pairs is avoided; see, e.g., Efron \& Tibshirani 1993, see also
the critical remarks on the use of the bootstrap method in point
process statistics given in Snethlage 2000) gives $1\sigma$ error
estimates of the order of 80 percent (cosmic variance not
included). The detection of the second maximum in the power spectrum
on such large scales, if real, would have very important implications
on current structure formation models. However, as pointed out in the
Introduction, measurements on such large scales are easily biased by
very small systematic errors of the survey detection model. We
postpone a detailed study of this very  questionable feature to a
subsequent paper. The present investigation concentrates more
conservatively on the range $0.013\le k\le 0.4\,h\,{\rm Mpc}^{-1}$
which is found to be free from {\it any} significant artifical
fluctuations (Sect.\,\ref{SEARCH}), which can be easily monitored by
the available N-body simulations, and which contains density waves
well sampled by the REFLEX clusters.

Individual spectra obtained with the three {\it flux-limited}
subsamples F300 to F500 are shown in Fig.\,\ref{FIG_PSA_ALL}, now
including the $1\sigma$ errors adapted from the N-body
simulations. Whereas the spectra obtained with F300 and F400 (upper
and middle panel) show a maximum at $k\approx 0.03\,h\,{\rm
Mpc}^{-1}$, the F500 data (lower panel) suggest only a flattening of
the spectral densities. Especially the power spectral density obtained
at the fundamental mode seems to indicate a still rising power
spectrum for smaller $k$ values. A similar effect is seen in the
simulations (see Fig.\,\ref{FIG_PK_TEST}) suggesting a statistically
not very significant but noticable leakage of fluctuation power
especially from the second to the first fundamental mode. The
reference to Fig.\,\ref{FIG_PSA_HUGE} reveals that the fundamental
mode of F500 is already part of the second probably not real maximum
in the power spectrum at $k\approx 0.01\,h\,{\rm Mpc}^{-1}$. Hence the
fundamental mode of F500 should not necessarily get the highest weight
in the evaluation of the maximum of the power spectrum on smaller
scales. We test the possibility that the location of the maximum
increases with volume but could not find any systematic effect (see
Sect.\,\ref{CHAR}).

The spectra shown in Fig.\,\ref{FIG_PSA_VOL} obtained with the {\it
volume-limited} subsamples L050 and L120 (upper and middle panel) show
a broad maximum at $k\approx 0.03\,h\,{\rm Mpc}^{-1}$. A weak
indication is found for a positive slope on larger scales. The second
maximum of the power spectrum seen in Fig.\,\ref{FIG_PSA_HUGE} is not
sampled by L050 and L120 because their sample volumes do not reach
such large scales. The Fourier volumes are therefore restricted in
both cases to $(400\,h\,{\rm Mpc}^{-1})^3$. For comparison the lower
panel shows the power spectrum obtained with the {\it flux-limited}
subsample F400 estimated within the same volume as used for L050 and
L120. In general, the overall shapes of the spectra obtained with the
volume- and flux-limited subsamples are found to be similar, although
minor differences might be seen on smaller scales (see below). The
three spectra also show that the amplitude increases with increasing
lower X-ray luminosity of the subsample. However, larger sample sizes
are needed to confirm the effect.

To summarize, basically all REFLEX spectra are consistent with a broad
maximum of the cluster power spectrum at comoving wavenumbers around
$k\approx 0.03\,h\,{\rm Mpc}^{-1}$ corresponding to wavelengths of
about $200\,h^{-1}\,{\rm Mpc}$. A second maximum is found at
$k=0.01\,h^{-1}\,{\rm Mpc}$ corresponding to $600\,h^{-1}\,{\rm Mpc}$,
but appears  questionable (see Sect.\,\ref{DISCUSS}). These
findings are summarized in Fig.\,\ref{FIG_POWER}, showing the combined
spectra obtained with the subsamples F300 to F500,  and
illustrating the stability of the results obtained within different
volumes. We regard this as a representative REFLEX power spectrum.

\subsection{First cosmological implications}\label{CHAR}
In the following we want to characterize the overall shape of the
observed power spectra as well as specific spectral features like the
location of the maximum and the local slope of $P(k)$ in specific $k$
ranges, restricting the discussion mainly to the conservative $k$
range $0.013\le k\le 0.4\,h\,{\rm Mpc}^{-1}$ mentioned above. This
will enable us to derive our first cosmological implications.

\begin{figure}
\vspace{-0.3cm}
\centerline{\hspace{2.8cm}
\psfig{figure=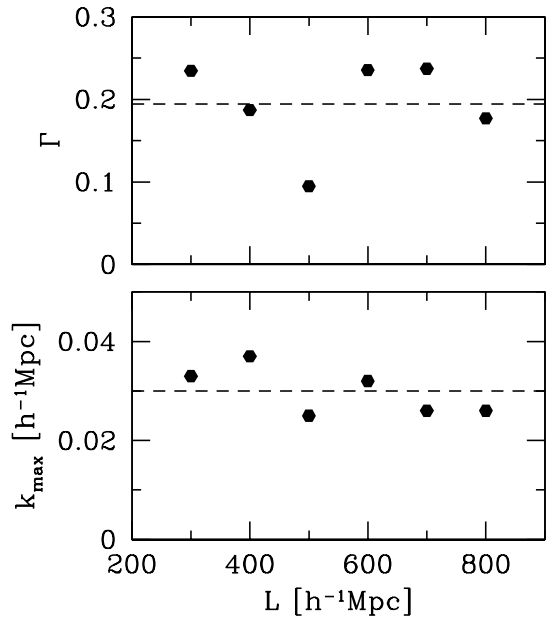,height=12.0cm,width=12.0cm}}
\vspace{-3.5cm}
\caption{\small Upper panel: values of the shape parameter, $\Gamma$,
as a function of the box length, $L$, of the Fourier volume, obtained
from fits of the linear CDM model to the flux limited subsamples F300 to
F800 (from left to right). Lower panel: values of the wavenumber of the
maximum of the power spectrum, $k_{\rm max}$, as a function of $L$
obtained from fits of the phenomenological model. The sample averages
are indicated by dashed lines.}
\label{FIG_FITTAB}
\end{figure}

\begin{figure}
\vspace{-0.5cm}
\centerline{\hspace{0.0cm}
\psfig{figure=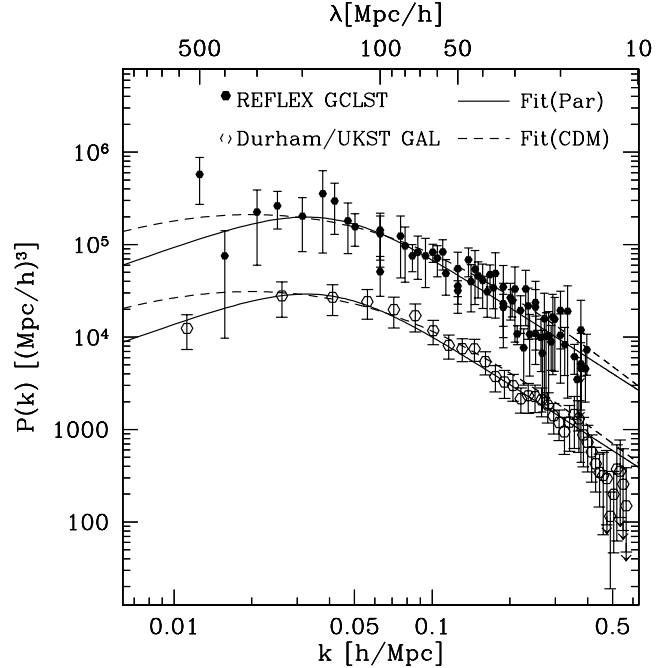,height=9.0cm,width=9.0cm}}
\vspace{-0.1cm}
\caption{\small Combined power spectrum obtained with the REFLEX
cluster subsamples F300 to F500 (filled symbols, measurements on
scales $<20\,h^{-1}\,{\rm Mpc}$ omitted) compared with the power
spectrum of Durham/UKST galaxies (open symbols) obtained by Hoyle et
al. (1999). The upper lines give the fit to the REFLEX cluster power
spectra, the lower lines the same fits divided by the squared `biasing
factor' $b^2=2.6^2$. The continuous line is a fit of the
phenomenological model (Par), the dashed line the CDM-like model fit
(CDM) as described in Fig.\,\ref{FIG_POWER}.}
\label{FIG_PSA_COMP_GAL}
\end{figure}

The REFLEX power spectral densities shown in the last sections are
sampled strictly following the rules of standard Fourier analysis. As
a consequence we have to work with uncomfortably large but
statistically almost independent $k$ bins which complicates the
analyses even of the maximum of $\hat{P}(k)$ in the conservative $k$
range. To improve the `eye ball' estimates of the location of the
maximum of the power spectra given in Sect.\,\ref{SHAPE} and to get a
handle of the expected errors, the spectra are parameterized in two
different ways. The first method applies a purely phenomenological
fitting function which gives an almost model-independent description
of the data (see also Peacock 1999, p.\,530):
\begin{equation}\label{PDFIT}
P(k)=A\,k_0^{-i_1}\,k^{i_1-3}\,\left[1+\left(\frac{k_{\rm c}}{k}\right)^{i_2}\right]^{-1}\,.
\end{equation}
The location of the maximum of the power spectrum is
\begin{equation}\label{PKMAX}
k_{\rm max}=k_{\rm c}\left(\frac{-n}{n_{\rm
s}}\right)^{\frac{1}{n-n_{\rm s}}}\,.
\end{equation}
The slope on large scales, $n=i_1+i_2-3$, is set to `1' because no
statistically reliable information is attainable from the REFLEX data
in this scale range. The slope on small scales is $n_{\rm
s}=i_1-3$. The characteristic scale, $k_{\rm max}$, is comparable to
the wavenumber corresponding to the horizon length at the epoch of
matter-radiation equality, $k_{\rm eq}=0.195\,\Omega_0\,h^2\,{\rm
Mpc}^{-1}$ (see Peebles 1993, p.\,164, and Peacock 1999, p.\,459), and
thus yields an estimate of the cosmic mass density (assuming that 3
relativistic neutrino families are left over from high redshift, and
that neutrino masses are small compared to the temperature of the
cosmic microwave background radiation),
\begin{equation}\label{OMEGA1}
\Omega_0\,h\,=\,5.13\,\frac{k_{\rm max}}{h\,{\rm Mpc}^{-1}}\,.
\end{equation}
In contrast to the first method which allows a variable slope at small
scales and a narrow maximum, the second method is less flexible, but
physically better defined. The fitting function is based on the CDM
linear transfer function, $T(k)$, as given in Bardeen et al. (1986),
where the power spectrum, $P(k)=AkT^2(k)$, again is assumed to have
$n=1$ on very large scales. The shape of the power spectrum is
characterized by the shape parameter, $\Gamma$, defined in the
standard way (eq.\,\ref{MOD2}), giving the approximate relation
between $\Gamma$ and $k_{\rm max}$
\begin{equation}\label{KMAX}
k_{\rm max}\,=\,0.114\,\Gamma\,h\,{\rm Mpc}^{-1}\,.
\end{equation}
For CDM-like models with low baryon density, $\Gamma$ is mainly
determined by $\Omega_0$, $h$, and $\Omega_{\rm b}$ (see
eq.\,\ref{MOD2}). This equation offers another way to approximate the
cosmic mass density via
\begin{equation}\label{OMEGA2}
\Omega_0\,h\,\approx\,\Gamma.
\end{equation}
A standard SIMPLEX $\chi^2$ minimization method is applied separately
to the spectra obtained with the subsamples F300 to F800 to perform
numerical fits from which the values of $k_{\rm max}$ and $\Gamma$ are
deduced. This assumes that the power spectral densities of each
individual spectrum are statistically independent. For the given
REFLEX survey window ($|W_k|^2\le 0.082$ for all $k$ and volumes
studied), and for the given spacing of the $k$ values of the measured
$P(k)$ data at the multiples of the fundamental mode this is
approximately the case. The values of $k_{\rm max}$ and $\Gamma$
obtained from the fits are independent of the volumes used to perform
the Fourier analyses as shown in Fig.\,\ref{FIG_FITTAB}, strongly
supporting the detection of a real maximum of $P(k)$ in the given $k$
range. Averages and their formal $1\sigma$ standard deviations of
$k_{\rm max}$ and $\Gamma$ using the subsamples F300 to F800 give for
the two fit functions, respectively,

\begin{equation}\label{PARAMVAL}
k_{\rm max}\,=\,0.030\pm0.005\,,\quad\quad\Gamma\,=\,0.195\pm0.055\,.
\end{equation}
The $\Gamma$ estimate corresponds to $k_{\rm max}=0.022\pm
0.006$. Similar numbers are obtained when only the subsamples F300 to
F500 are used. Note that the subsamples F300 to F800 are statistically
dependent so that the error estimates given in (\ref{PARAMVAL}) must
be regarded as lower limits. The phenomenological and the CDM-like
model based on these mean values are shown as continuous and dashed
lines, respectively, in the Figs.\,\ref{FIG_PSA_HUGE},
\ref{FIG_PSA_ALL}, \ref{FIG_PSA_VOL},
\ref{FIG_POWER}, and \ref{FIG_PSA_COMP_GAL}. They both give a good
description of the {\it shape} of all power spectra obtained with the
flux- and with the volume-limited REFLEX subsamples. In
Fig.\,\ref{FIG_POWER} we show this for the combined power spectrum
obtained with the flux-limited subsamples F300 to F500. The two
methods give consistent results for the location of the maximum of the
REFLEX power spectra in the range
\begin{equation}\label{KMAXRANGE}
0.022\pm 0.006\,\le\,k_{\rm max}\,\le\,0.030\pm 0.005\,h\,{\rm Mpc}^{-1}\,.
\end{equation}
Similarily, the values of the cosmic density parameter obtained with
eq.\,(\ref{OMEGA1}) and $k_{\rm max}$ from (\ref{PARAMVAL}), and with
eq.\,(\ref{OMEGA2}) and $\Gamma$ from (\ref{PARAMVAL}) give the range
\begin{equation}\label{OMEGARANGE}
0.15\pm 0.03\,\le\,\Omega_0\,h\,\le\,0.20\pm 0.06\,.
\end{equation}

It is interesting to note that the spectral slopes on {\it small}
scales of the volume-limited subsamples L050 and L120 estimated by fitting
(\ref{PDFIT}) are found to be slightly flatter, $n_{\rm s}=-1.6\pm
0.4$, compared to $n_{\rm s}=-1.8\pm 0.4$ obtained with the
flux-limited subsamples ($1\sigma$ error estimates from N-body
simulations). However, the differences are statistically perhaps not
very significant and might be attributed to cosmic variance.

\begin{figure}
\vspace{-1.0cm}
\centerline{\hspace{0.0cm}
\psfig{figure=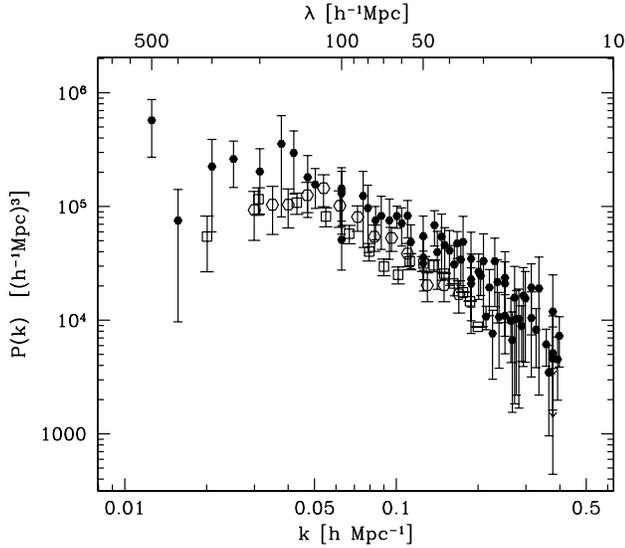,height=9.0cm,width=9.0cm}}
\vspace{-0.8cm}
\caption{\small Combined REFLEX power spectrum obtained with the
subsamples F300 to F500 (filled symbols) compared to the power
spectrum obtained from Abell/ACO clusters (open hexagons) by Retzlaff
et al. (1998) and from APM clusters (open squares) by Tadros et
al. (1998).}
\label{FIG_PSA_COMP_OBS}
\end{figure}

Figure\,\ref{FIG_PSA_COMP_OBS} compares the REFLEX power spectrum with
the Abell/ACO (Retzlaff et al. 1998, see also Einasto et al. 1997) and
the APM (Tadros et al. 1998) spectra. The respective amplitudes of the
power spectra of the Abell/ACO and APM samples are 1.7 and 2.2 below
REFLEX. This might be attributed to the different cluster luminosities
contained in the samples. For $k\ge 0.08\,h\,{\rm Mpc}^{-1}$ the
spectra give consistent slopes of approximately $-1.8$ although both
the REFLEX and the Abell/ACO sample do not show the minimum at
$k\approx 0.1\,h\,{\rm Mpc}^{-1}$ found with the APM sample. Regarding
the maximum of $P(k)$ the Abell/ACO data suggest a comparatively
narrow peak at $k_{\rm max}=0.05\,h\,{\rm Mpc}^{-1}$ consistent with
the estimate of Einasto et al. (1997). Contrary to this the REFLEX
spectrum has a broad maximum which peaks in the range $0.022\le k_{\rm
max}\le 0.030\,h\,{\rm Mpc}^{-1}$. Note that the exact evaluation of
the statistical significance of this difference is difficult to assess
because the REFLEX and Abell/ACO power spectra are sampled in
different ways. The broad maximum of the REFLEX spectrum appears to be
more consistent with the APM sample if the REFLEX measurement at
$500\,h^{-1}\,{\rm Mpc}$ is excluded.

Figure\,\ref{FIG_PSA_COMP_GAL} compares the combined REFLEX power
spectrum obtained with the flux-limited subsamples F300 to F500 with
the spectrum obtained with a magnitude-limited sample of Durham/UKST
galaxies (Hoyle et al. 1999). We chose this sample because of the
comparatively large samples size (2501 galaxies, 1 in 3 sampling
rate), the large volume (1\,450 square degrees, $z\le 0.1$), and the
small effects of the survey window. Recall that the upper continuous
line is the fit of the phenomenological model to the REFLEX data, the
upper dashed line the fit of the CDM-like model; the lower lines are
the same fits shifted by the factor $6.8$. For wavelengths
$20<\lambda<300\,h^{-1}\,{\rm Mpc}$ the overall shapes of the cluster
and galaxy power spectra are very similar. The ratio of the linear
biasing factors for the given REFLEX cluster subsample and the galaxy
sample as deduced from the shift factor is $b=2.6$.
\section{Comparison with CDM models}\label{MODELS}
\subsection{Semi-analytic model}\label{SEMOD}
To make a first comparison with cosmological models and an attempt to
differentiate between their presently discussed variants, an outline
of a semi-analytic model is given for biased nonlinear power spectra
in redshift space for clusters of galaxies. The model gives a good
overview of the effects of different model parameters and is used to
narrow the parameter ranges needed for a more detailed comparison with
N-body simulations. Notice that a significant number of N-body
simulations has to be performed for each parameter set in order to
derive statistical meaningful error estimates which is  planned
for the second paper on the REFLEX power spectrum. The model spectra
are computed with parameter values taken from the literature and are
compared with the REFLEX power spectra. No evolution of structures is
assumed within the redshift range covered by the REFLEX subsamples
analyzed ($z<0.15$, for an exact treatment see also Magira et
al. 2000). The linear power spectrum $P(k)\sim k^n\,T^2(k)$ is
normalized by the standard deviation of the density contrast,
$\sigma_8$, obtained with the spherical top hat filter function with
the filter radius $R=8\,h^{-1}\,{\rm Mpc}$. The fitting formula for
the linear transfer function, $T(k)$, from Bardeen et al. (1986) is
applied with the shape parameter (Sugiyama 1995)
\begin{equation}\label{MOD2}
\Gamma\,=\,\Omega_0\,h\,\exp \left(-\Omega_{\rm
b}-\sqrt{\frac{h}{0.5}}\frac{\Omega_{\rm b}}{\Omega_0}\right)\,,
\end{equation}
appropriate for models with low present baryon densities, $\Omega_{\rm
b}$. The mapping between evolved and non-evolved real-space power
spectra can be deduced from fits to the results obtained with N-body
simulations (Hamilton et al. 1991, Peacock \& Dodds 1994). We apply
the prescription in the form presented in Mo, Jing \& B\"orner (1997,
and references given therein). The approximation of the linear growth
factor at redshift zero is taken from Carroll, Press \& Turner
(1992). It was shown that this mapping is almost independent of the
slope of the linear power spectrum. Moreover, as long as the strongly
non-linear clustering regime is excluded the formalism reproduces the
power spectra obtained from N-body simulations quite well.

To compute the observed or effective biasing values Moscardini et
al. (2000, see also Matarrese et al. 1997 and Borgani et al. 1999)
assumed a linear biasing between matter and object number density
fluctuations, a reasonable assumption in the linear regime. They
derived an exact relation between the observed and the matter
two-point spatial correlation function (their equation\,7) which we
reproduce in $k$-space, ignoring any redshift-dependence of the
correlation function. The real-space (evolved) power spectrum thus
reads
\begin{eqnarray}\label{MOD15}
P_{\rm
real}(k)\,=\,P(k)\,\left[\int_Z\,dz_1\,N(z_1)r^{-1}(z_1)\right]^{-2}\hspace{2.0cm}\\
\nonumber \int_Z\,dz_1\,dz_2\,N(z_1)\,r^{-1}(z_1)\,b_{\rm
eff}(z_1)\,b_{\rm eff}(z_2)\,r^{-1}(z_2)\,N(z_2)\,.
\end{eqnarray}
where $N(z)dz$ is the number of clusters expected in the redshift
interval $[z,z+dz]$,
\begin{equation}\label{MOD16}
b_{\rm eff}(z)\,=\,\frac{1}{N(z)}\,\int_M\,dM\,b(M)\,N(M,z)\,,
\end{equation}
the effective biasing at redshift $z$, $N(M,z)dM\,dz$ the number of
clusters expected in the mass range $[M,M+dM]$ and in the redshift
range $[z,z+dz]$, and $r(z)$ the comoving distance of the redshift
shell $z$. Note that $N(M,z)dM=\tilde{N}(L,z)dL$ so that (\ref{MOD16})
can also be used for luminosities once the $M-L$ conversion is
performed (see below). We evaluate the integrals at the redshifts and
luminosities given by the observed subsample for which the effective
biasing factor should be determined (for each cluster we compute from
the luminosity the mass and the corresponding biasing factor and plug
the result into the two equations given above). This circumvents the
introduction of additional Press-Schechter -like models for the cosmic
mass function which we do not intend to test in the present context
(probably introducing some inconsistency between the observed and the
model luminosity function), but guarantees a one-to-one correspondence
between the clusters used to measure the power spectrum and the
clusters used to estimate the effective biasing for the subsample
under consideration. This, however, increases the variances but
reduces the systematic errors to a minimum.

For clusters of galaxies simple biasing schemes are expected
(Sect.\,\ref{INTRO}). In this respect the model of Mo \& White (1996)
is of special interest. They combine (a) conditional probability
densities derived by Bond et al. (1991) for Gaussian random fields
within the general framework of Markovian diffusion processes with an
`absorbing barrier' at the critical density contrast, with (b)
gravitationally induced motions as predicted by a spherical 
collapse model. We use the fitting formula given in Sheth \& Tormen
(1999) which is found to give a better agreement with N-body on small
scales. The critical overdensity, $\delta_{\rm c}$, is determined by
the cosmological background model as described in Kitayama \& Suto
(1996). The relation between mass and radius is
$M\,=\,\frac{\Omega_0\,H_0^2\,R^3}{2\,G}$.

For the $M-L_{\rm X}$ conversion the empirical relation between the
total mass $M$ and X-ray luminosity within $r_{200}$ is used (Reiprich
\& B\"ohringer 2000):
\begin{equation}\label{MLR200}
\frac{M}{h^{-1}\,M_\odot}\,=\,4.7\times 10^{14}\,
\left(\frac{L_{\rm X}}{10^{44}\,h^{-2}\,{\rm erg}\,{\rm s}^{-1}}\right)^{\frac{1}{1.243}}\,.
\end{equation}
The systematic effects caused by using the relation obtained with
$r_{500}$ are investigated below (Fig.\,\ref{FIG_MODEL_TEST}).  We assume
that the masses deduced from (\ref{MLR200}) closely resemble the
virial masses.

The transformation of the real-space power spectrum into redshift
space is determined by the effects of peculiar velocities and redshift
measurement errors. If the maximum distances are large compared to
$k^{-1}$ (distant observer approximation) only the linear flow of the
velocity field makes an additional contribution to the fluctuation
field in redshift space (Kaiser 1987). On small scales the peculiar
velocities and the redshift measurement errors of the clusters smooth
the fluctuation field which can be described by a Lorentzian
distribution in $k$-space. The two effects can be integrated over the
cosine, $\mu$, of the angle between the normal vector of the density
wave in $k$-space and the line-of-sight, and give
\begin{equation}\label{MOD24}
P_{\rm obs}(k)\,=\,P_{\rm
real}(k)\,\int_0^1\,d\mu\,\frac{(1+\beta\mu^2)^2}
{(1+k^2\mu^2\frac{\sigma^2}{2H^2})^2}\,,
\end{equation}
\begin{equation}\label{MOD241}
\sigma\,=\,\sqrt{\frac{\sigma^2_{\rm P}}{2}+\sigma_z^2}\,,\quad\quad
\beta\,=\,\frac{\Omega_0^{0.6}}{b_{\rm eff}}\,.
\end{equation}
Here, $\sigma_{\rm P}$ is the pairwise velocity dispersion of the dark
matter haloes and $\sigma_z$ the average error of the cluster
redshifts. The effects of $\Omega_\Lambda$ are not important (see
Lahav et al. 1991). For the computation of $\sigma_{\rm P}$ we
concentrate on the effects on scales (pair separations) larger
$5\,h^{-1}\,{\rm Mpc}$ and neglect correlated motions so that
$\sigma_{\rm P}$ can be approximated by (Peebles 1980, Sect.\,72)
\begin{equation}\label{PEC1}
\sigma^2_{\rm P}\,=\,\frac{2}{3}\,<v_1^2>\,.
\end{equation}
It is thus expected that the cluster motion inherit a random velocity
from the random motion of the overall matter distribution. The
density-weighted mean square peculiar velocity is determined by the
integral of the cosmic energy equation as derived in the BBGKY
hierarchy (e.g., Peebles 1980, Sect.\,74, Mo et al. 1997):
\begin{equation}\label{PEC2}
<v_1^2>\,=\,\frac{3}{2}\Omega_0\,H^2_0\,
\int_0^\infty \frac{dk}{k^3}\,\Delta_{\rm E}^2(k)\,.
\end{equation}
The cluster redshift errors can be deduced from the redshift errors of
the individual galaxies ($\sigma_{\rm g}=200\,{\rm km}\,{\rm s}^{-1}$,
Guzzo et al., in preparation), the number of cluster galaxies used to
estimate the cluster redshift ($N_z=5$) and the line-of-sight cluster
velocity dispersion ($\sigma_{\rm LOS}=700\,{\rm km}\,{\rm s}^{-1}$,
see Zabludoff et al. 1993). Although the latter quantity depends on
the specific structure formation model we take the empirical estimate
because in the present case we are only interested in a redshift error
estimate where the effects of the structure formation models are of
second-order. Following Danese, de\,Zotti \& di\,Tullio (1980) the
squared error of the cluster redshift is
\begin{equation}\label{PEC3}
\sigma_z^2\,=\,\frac{1}{N_z}\,\left[1.178\,\sigma_{\rm
LOS}^2+\sigma_{\rm g}^2\right]\,,
\end{equation}
where the factor 1.178 results from the 68 percent confidence interval
of the Student's t-distribution with 4 degrees of freedom (for
$N_z=4,3,2$ we computed the factors 1.242, 1.391 and 2.057,
respectively). Typical redshift errors of the REFLEX clusters of
galaxies are thus expected to be $\sigma_z=350\,{\rm km}\,{\rm
s}^{-1}$.
\subsection{Test of the model}\label{TEST}
\begin{figure}
\vspace{-1.8cm}
\centerline{\hspace{0.0cm}
\psfig{figure=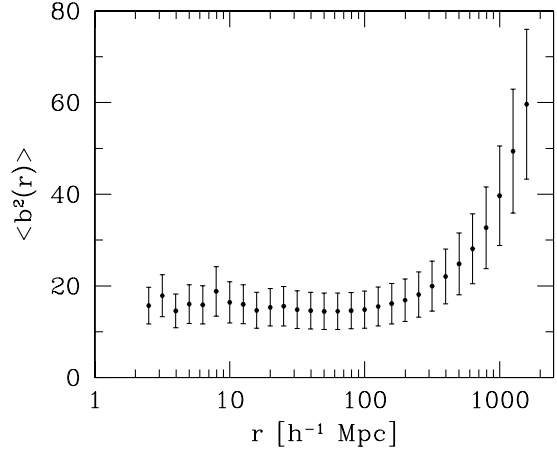,height=8.0cm,width=8.0cm}}
\vspace{-0.4cm}
\caption{\small Squared biasing factors computed with
eq.\,(\ref{BIASR}) for the F0 subsample as a function of pair
separation $r$. The $1\sigma$ error bars include the effects of the
errors introduced by the 50 percent uncertainty of the X-ray
mass-luminosity relation.}
\label{FIG_BIASR}
\end{figure}
\begin{figure}
\vspace{-1.0cm}
\centerline{\hspace{0.0cm}
\psfig{figure=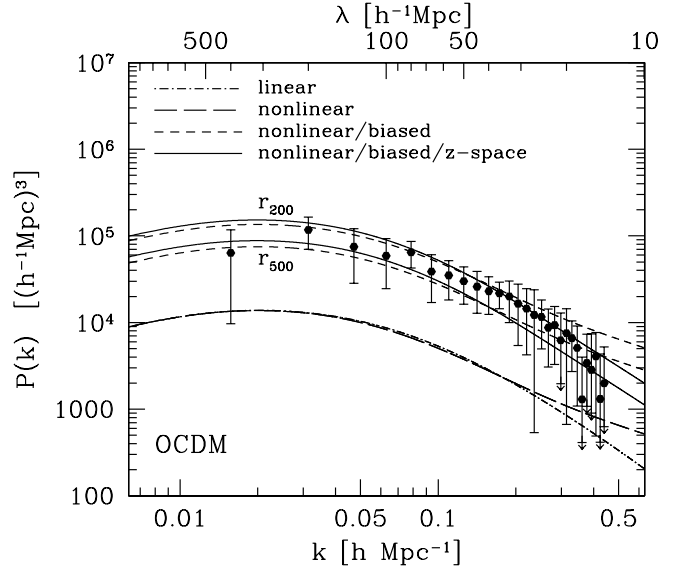,height=9.0cm,width=9.0cm}}
\vspace{-0.7cm}
\caption{\small Test of the semi-analytic model with N-body
simulations. Shown is the power spectrum averaged over 10 OCDM N-body
simulations (filled symbols) of ideal clusters samples
(Sect.\,\ref{SIMUL}) and their $1\sigma$ standard deviations (error
bars). The lines represent the power spectra obtained with the
semi-analytic model for the same model parameters as the N-body
simulations ($\Omega_0=0.40$, $\Omega_\Lambda=0$, $\Omega_{\rm
b}=0.05$, $h=0.60$, $\Gamma=0.20$, $n=1$, $\sigma_8=0.80$, i.e.,
cluster-normalized). Dashed-dotted line: linear matter power
spectrum. Long-dashed line: evolved matter power
spectrum. Short-dashed lines: evolved power spectrum including
effective biasing. Continuous lines: last spectrum transformed into
redshift space. The two types of spectra are shown for the
mass-luminosity relation obtained with $r_{200}$ and $r_{500}$.}
\label{FIG_MODEL_TEST}
\end{figure}
We test (\ref{PEC1}) and (\ref{PEC2}) with the available OCDM N-body
simulations (parameters are given in Tab.\,3), yielding the mean
cluster peculiar velocity $\sqrt{<v_1^2>}=516\,{\rm km}\,{\rm s}^{-1}$
(the standard deviation of this quantity for different simulations is
3 percent), and for pair separations $>10\,h^{-1}\,{\rm Mpc}$ the
approximately constant (within about 3 pecent) pairwise cluster
velocity dispersion $\sigma_{\rm P}=430\,{\rm km}\,{\rm
s}^{-1}$. Within the given errors the relation between these
quantities is reproduced by (\ref{PEC1}).  The simulations give a
maximum at $3\,h^{-1}\,{\rm Mpc}$ of $\sigma_{\rm P}=500\,{\rm
km}\,{\rm s}^{-1}$ and values of about $200\,{\rm km}\,{\rm s}^{-1}$
on smaller scales. The semi-analytic model neglects the small-scale
dependency because the REFLEX power spectrum does not sample the
corresponding $k$ range. On the other hand using (\ref{PEC1}) and
(\ref{PEC2}) the semi-analytic model predicts $\sigma_{\rm
P}=360\,{\rm km}\,{\rm s}^{-1}$, which is about 15 percent too small
compared to the simulations. We found this approximation good enough
for the real-redshift space transformation.

An important assumption implicitely used for the derivation of
(\ref{MOD15}) is that the averaged biasing factor is independent of
pair separation, $r$. For flux-limited samples one might expect that
at large $r$ the fraction of pairs consisting preferentially of at
least 1 luminous cluster could artifically increase the effective
biasing factor. This would increase the measured power spectral
densities at small $k$ and thus steepen the slope compared to the
volume-limited case. To test this, the number of pairs with separation
$r$ are weighted with the individual biasing factors of the pair
members, yielding the average squared biasing factors,
\begin{equation}\label{BIASR}
<b^2(r)>\,=\, \frac{ \sum_{\{(i,j)|r\le|\vec{r}_i-\vec{r}_j|<r+\Delta
r\}}\,b(M_i)\,\,b(M_j)} {N_{\rm P}(r)}\,,
\end{equation}
where $N_{\rm P}(r)$ is the number of cluster pairs with separations
within $[r,r+\Delta r]$. The mass variances, $\sigma^2(M)$, used for
the computation of $b(M_i)$ are derived assuming a scale-invariant
power spectrum with the spectral index $-2$.  Fig.\,\ref{FIG_BIASR}
shows the average squared biasing factors as a function of pair
separation for the REFLEX subsample F0. For pair separations
$r<150\,h^{-1}\,{\rm Mpc}$ no scale-dependent correlations between the
individual biasing factors are seen. For larger separations the
systematic increase of the average squared biasing factor suggests
that the treatment of effective biasing as described above must be
modified. The maximum pair separation, $r$, corresponds to the minimum
wavenumber $k\approx \pi/2r=0.010\,h\,{\rm Mpc}^{-1}$. For wavenumbers
larger than this limit no systematic errors are expected.  It will be
seen that the observed REFLEX power spectra which are compared with
the biasing model do not reach this limit. Note that this refers only
to REFLEX and must be re-evaluated for other surveys.

The semi-analytic model is tested against the 10 N-body simulations
(OCDM) of ideal cluster samples described in Sect.\,\ref{SIMUL}. In
Fig.\,\ref{FIG_MODEL_TEST} the lines give the theoretical spectra
obtained under the different model assumptions, the filled symbols the
average power spectral densities obtained from the N-body simulations,
and the error bars their $1\sigma$ standard deviations. The overall
agreement between model and simulation is good enough to separate
between different scenarios of cosmic structure formation. The largest
ambiguity is introduced by the specific choice of the mass-luminosity
relation. In the following the theoretical spectra obtained with
$r_{200}$ are shown because the corresponding cluster masses are
expected to give better estimates of the virial masses.

\begin{table}
{\bf Tab.\,3.} Model parameters of CDM variants used for the
semi-analytic model.
\label{T_MODEL}
\[
\hspace{1.2cm}\begin{array}{lcccccccll}
\hline
\noalign{\smallskip}
{\rm Model} & \Omega_0 & \Omega_\Lambda & h    & n   & \Omega_{\rm b} & \Gamma &
\sigma_8 \\
\noalign{\smallskip}
\hline 
{\rm SCDM}            & 1.00 & 0.00 & 0.50 & 1.0 & 0.050 & 0.45 &
1.37 \\
{\rm OCDM}            & 0.40 & 0.00 & 0.60 & 1.0 & 0.050 & 0.20 &
0.80 \\
\Lambda{\rm CDM}    & 0.30 & 0.70 & 0.65 & 1.0 & 0.036 & 0.21 & 
0.93 \\
{\rm TCDM}            & 1.00 & 0.00 & 0.50 & 0.8 & 0.100 & 0.41 &
0.58 \\
\tau{\rm CDM}       & 1.00 & 0.00 & 0.50 & 1.0 & 0.050 & 0.21 & 
0.60 \\
\noalign{\smallskip}
\hline
\end{array}
\]
\end{table}

\subsection{Results}\label{RESULTS}
\begin{figure}
\vspace{-1.0cm}
\centerline{\hspace{0.0cm}
\psfig{figure=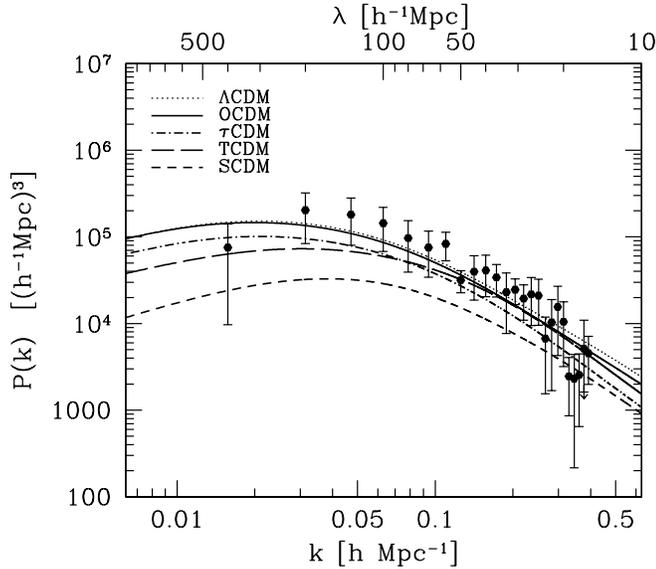,height=9.0cm,width=9.0cm}}
\vspace{-0.7cm}
\caption{\small Comparison of observed power spectral densities and
expectations of variants of CDM semi-analytic models for the
flux-limited subsample F400. The model parameters are given in
Tab.\,2.}
\label{FIG_W}
\end{figure}

As an example, in Fig.\,\ref{FIG_W} the REFLEX power spectrum obtained
with the F400 subsample is compared with different variants of CDM
models (the data obtained with F300 and F500 give similar
results). The values of the model parameters are given in Tab.\,3. The
standard Cold Dark Matter (SCDM) model with the COBE normalization as
given in Bennett et al. (1994) is shown for reference. The open CDM
(OCDM) model is cluster-normalized (see Sect.\,\ref{SIMUL}). For the
low-density flat ($\Lambda$CDM) model see Liddle et al. (1996a,b). The
tilted (TCDM) model is described in Moscardini et al. (2000) and the
references given therein. The $\tau$CDM model is cluster-normalized
according to Viana \& Liddle (1996).

The measured power spectra discriminate between the models, SCDM and
TCDM are excluded, $\tau$CDM fits marginal the lower $1\sigma$ range,
the open and $\Lambda$CDM models slightly underpredict the fluctuation
amplitude but within the $1\sigma$ significance range.

\begin{figure}
\vspace{-1.0cm}
\centerline{\hspace{0.0cm}
\psfig{figure=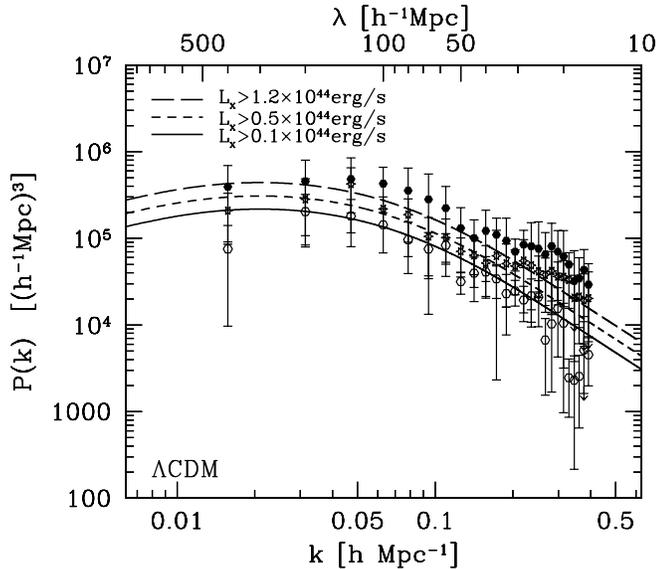,height=9.0cm,width=9.0cm}}
\vspace{0.0cm}
\caption{\small Comparison of observed power spectral densities and
predictions of the $\Lambda$CDM semi-analytic model for the
flux-limited subsample F400 (open hexagons) and for the volume-limited
subsamples: stars for $L_{\rm X}>0.5\times 10^{44}\,{\rm erg}\,{\rm
s}^{-1}$ (subsample L050), filled symbols for $L_{\rm X}>1.2\times
10^{44}\,{\rm erg}\,{\rm s}^{-1}$ (subsample L120), $h=0.5$. The
$\Lambda$ model is renormalized to $\sigma_8=0.70$ to give a good fit
to the flux-limited subsample.}
\label{FIG_B}
\end{figure}

To test the biasing trends we changed the $\Lambda$CDM normalization
from $\sigma_8=0.93$ to $\sigma_8=0.70$ (similarly we could also
change $\sigma_8=0.80$ to $\sigma_8=0.60$ for the OCDM model) yielding
an acceptable fit to the flux-limited REFLEX power spectrum (open
symbols and continuous line in Fig.\,\ref{FIG_B}). The $\Lambda$CDM
spectra are then computed for the same volume-limited subsamples as
used for the determination of the empirical spectra. The increase of
the amplitude with the increasing lower X-ray luminosity -- although
at the detection limit of REFLEX -- is well reproduced by the model,
but not the apparent flattening of the slope on scales
$<100\,h^{-1}\,{\rm Mpc}$. However, the errors of the slope
measurements as deduced from the simulations are quite large so that
the apparent difference might not be statistically
significant. Moreover, neither the scale-independency of the effective
biasing parameter in this range (see Fig.\,\ref{FIG_BIASR}) nor the
analyses of the OCDM simulations suggest such an effect.

\section{Discussion and conclusions}\label{DISCUSS}
The most important result of the present investigation is the
detection of a broad maximum of the power spectrum of the fluctuations
of comoving number density of X-ray selected cluster galaxies in the
range $0.022\le k\le 0.030\,h\,{\rm Mpc}^{-1}$
(Fig.\,\ref{FIG_POWER}). The maximum is flatter and peaks at a smaller
wavenumber compared to optically selected cluster samples.  On scales
$0.02\le k\le 0.4\,h\,{\rm Mpc}^{-1}$ the similiarity to the spectra
obtained from optically selected galaxy samples is striking
(Fig.\,\ref{FIG_PSA_COMP_GAL}). In this range the REFLEX data rule out
galaxy formation models with strongly nonlinear biasing schemes.

Within the course of the exploration of the REFLEX data and the
results of the N-body simulations we found that for surveys smaller
than REFLEX cosmic variance might be more important than previous
studies suggest. For example, the variation of the comoving cluster
number density along the redshift direction shows a huge underdense
region located between $z\approx 0.015$ and 0.045 in the southern
hemisphere where the comoving cluster number density drops by a factor
of 3 below the mean level (Fig.\,\ref{FIG_CNZ}). This complicates the
determination of the local cluster luminosity function, at least for
the less rich systems (B\"ohringer et al., in preparation). Another
example is the variation of the linear slope of the cumulative
flux-cluster number counts between $-1.6$ and $-1.2$ as found in the
N-body simulations (Fig.\,\ref{FIG_SHISTO}). {\it We regard this as a
warning not to draw general cosmological conclusions from cluster
samples with a size smaller than REFLEX}.

The REFLEX data show extra fluctuation power on scales $k\approx
0.01\,h\,{\rm Mpc}^{-1}$ (Fig.\,\ref{FIG_PSA_HUGE}). From our
simulations we found that artifical power spectral densities of an
order of magnitude can be easily produced on $500\,h^{-1}\,{\rm Mpc}$
scales if, e.g., the lower X-ray luminosity limit of $L_{\rm X}^{\rm
min}=1.0\times 10^{43}\,{\rm erg}\,{\rm s}^{-1}$, which is used in the
present investigation to get almost complete REFLEX subsamples, would
be erroneously underestimated by a factor of about 1.5. Similarily,
already on scales of $400\,h^{-1}\,{\rm Mpc}$ small changes in the
method to estimate the radial part of the selection function (compare
the results obtained with smoothed redshift distributions and X-ray
luminosity functions, Fig.\,\ref{FIG_COMP_XLF_SMOOTH}) change the
power spectral densities by a factor 1.6. These two examples
illustrate the difficulty measuring fluctuations on scales
$>400\,h^{-1}\,{\rm Mpc}$ which is the basic motivation for
restricting the present investigations more conservatively to the
small wavelength range.

Extra fluctuation power on $800\,h^{-1}\,{\rm Mpc}$ scales is also
found for the Abell/ACO richness $\ge 1$ clusters by Miller \& Batuski
(2000). In addition to the fact that they oversample the cluster power
spectrum which mimic a more significant effect than the data can
provide, it is difficult to understand how gradients in comoving
cluster number density by a factor of 2, corrected with crude
step-like radial selection functions, and the neglection of any
corrections for galactic extinction can lead to precise fluctuation
measurements at $800\,h^{-1}\,{\rm Mpc}$. It is surely insufficient to
use cluster quadrant counts showing a scatter of 16 percent to justify
fluctuation measurements aiming to detect fluctuations below the 1
percent level.

The REFLEX power spectra do not show any indication for a narrow peak
at $k=0.05\,h\,{\rm Mpc}^{-1}$.  The report of such a feature in the
power spectrum of Abell/ACO clusters and the interpretation as
evidence for a regular distribution of galaxy clusters with a
periodicity of $120\,h^{-1}\,{\rm Mpc}$ by Einasto et al.\ (1997)
implies substantial difficulties for current models of structure
formation.  Retzlaff et al.\ (1998) who have found a similar but less
peaked feature in the Abell/ACO cluster $P(k)$ used a large set of
N-body simulations to demonstrate the potential importance of cosmic
variance in this context. The discrepancy between REFLEX and Abell/ACO
cluster results might be attributed to the additional 35 percent
non-Abell/ACO/Supplement clusters included in the REFLEX
catalogue. Unfortunately, the subtle selection effects imposed by
optical cluster selection (Sect.\,\ref{INTRO}) makes a quantitative
discussion of this point almost impossible. In any case, due to
current sample depths, cluster power spectrum analyses are restricted
in general to volumes $<(500\,h^{-1}\,{\rm Mpc})^3$, and this imposes
a spectral resolution $\Delta k=0.013$ (fundamental mode) at best.
Therefore, a significant detection of a feature such as a peak of
width $\Delta k \approx 0.02$ is arguable at all.

The REFLEX spectra are compared with semi-analytic models describing
the biased nonlinear power spectrum in redshift space. Most of the
equations applied are calibrated with N-body simulations. We found
that structure formation models with a low cosmic mass density (OCDM,
$\Lambda$CDM) give the best representation of the REFLEX data
(Fig.\,\ref{FIG_W}). Although the models could reproduce the observed
changes of the amplitudes with samples of different luminosities, we
regard the results are tendatively. Larger sample sizes are necessary
to confirm this finding.

\begin{acknowledgements}                                                        

We thank Joachim Tr\"umper and the ROSAT team for providing the RASS
data fields and the EXSAS software, Harvey MacGillivray for providing
the COSMOS galaxy catalogue, Rudolf D\"ummler, Harald Ebeling,
Alastair Edge, Andrew Fabian, Herbert Gursky, Silvano Molendi,
Marguerite Pierre, Giampaolo Vettolani, Waltraut Seitter, and Gianni
Zamorani for their help in the optical follow-up observations at ESO
and for their work in the early phase of the project, Kathy Romer for
providing some unpublished redshifts, Sabino Matarrese for some
interesting discussions, and Stefano Borgani for critical reading of
the manuscript.  P.S. acknowledges the support by the Verbundforschung
under the grant No.\,50\,OR\,9708\,35, H.B. the Verbundforschung under
the grand No.\,50\,OR\,93065.

\end{acknowledgements}

\end{document}